\documentclass[12pt,english]{article}

\usepackage[longnamesfirst]{natbib}
\usepackage{url}
\makeatletter
\def\url@leostyle{%
    \def\UrlFont{\sf}}{\def\UrlFont{\small\ttfamily}}
\makeatother
\usepackage{geometry}
\usepackage[font=small,labelfont=bf]{caption}

\usepackage[hidelinks,breaklinks]{hyperref}
\usepackage[hyphenbreaks]{breakurl}

\usepackage{amsmath}

\usepackage{amsthm}

\usepackage{amsfonts}
\usepackage{dsfont}

\usepackage[T1]{fontenc}
\usepackage{txfonts} 

\usepackage{setspace}

\usepackage{microtype}

\usepackage[normalem]{ulem}

\usepackage{color}
\definecolor{darkgreen}{RGB}{0,128,20}
\definecolor{violet}{RGB}{143,0,255}

\usepackage{bbold}

\usepackage{enumitem}

\usepackage{graphicx}

\usepackage[english]{babel}

\numberwithin{equation}{section}

\usepackage[]{datetime2}

\usepackage[toc,page]{appendix}

\title{\textsc{The Open Systems View}}

\author{Michael E. Cuffaro\footnote{Munich Center for Mathematical Philosophy, LMU Munich, E-mail: mike@michaelcuffaro.com} \and Stephan Hartmann\footnote{Munich Center for Mathematical Philosophy, LMU Munich, E-mail: S.Hartmann@lmu.de}}

\newcommand{\beq}{\begin{equation}}
\newcommand{\eeq}{\end{equation}}
\newcommand{\bdm}{\begin{displaymath}}
\newcommand{\edm}{\end{displaymath}}
\newcommand{\bea}{\begin{eqnarray}}
\newcommand{\eea}{\end{eqnarray}}

\newcommand{\benum}{\begin{enumerate}}
\newcommand{\eenum}{\end{enumerate}}
\newcommand{\bit}{\begin{itemize}}
\newcommand{\eit}{\end{itemize}}
\newcommand{\bdes}{\begin{description}}
\newcommand{\edes}{\end{description}}
\newcommand{\bpic}{\begin{picture}}
\newcommand{\epic}{\end{picture}}
\newcommand{\bc}{\begin{center}}
\newcommand{\ec}{\end{center}}

\newcommand{\bq}{\begin{quote}}
\newcommand{\eq}{\end{quote}}

\shortcites{schlamminger2008}

\begin{document}

\singlespacing

\maketitle

\thispagestyle{empty}


\begin{abstract}
  There is a deeply entrenched view in philosophy and physics, the \emph{closed systems view}, according to which isolated systems are conceived of as fundamental. On this view, when a system is under the influence of its environment this is described in terms of a coupling between it and a separate system which taken together are isolated. We argue against this view, and in favor of the alternative \emph{open systems view}, for which systems interacting with their environment are conceived of as fundamental, and the environment's influence is represented via the dynamical equations that govern the system's evolution. Taking quantum theories of closed and open systems as our case study, and considering three alternative notions of fundamentality: (i)~ontic fundamentality, (ii)~epistemic fundamentality, and (iii)~explanatory fundamentality, we argue that the open systems view is fundamental, and that this has important implications for the philosophy of physics, the philosophy of science, and for metaphysics.
\end{abstract}

\section{Introduction}
\label{s:intro}

There is a deeply entrenched view in philosophy and physics according to which closed systems, i.e., systems not interacting with other systems, are conceived of as fundamental. For Gottfried Wilhelm Leibniz, for instance, the cosmos as a whole is an example of such a system; God forms it and sets it in motion, but then stands outside of it and allows it to proceed ever after in accordance with its own internal principles:

\begin{quote}
[T]he same force and vigor always remains in the world and only passes from one part to another in agreement with the laws of nature and the beautiful pre-established order. \citep[Leibniz's first letter, sec.\ 4]{leibnizClarke2000}
\end{quote}

In the context of physics, a system is considered to be closed if it does not exchange energy, matter, heat, information or anything else with its environment,\footnote{A system may also be closed with respect to only one such quantity. For instance, a system is \emph{chemically closed} if it does not exchange matter with the environment (though it might, for example, exchange heat with it).} and there are reasons stemming from physics to believe that the cosmos is the \emph{only} closed system that truly exists. For instance, it is well-known that gravity (unlike the, in many respects structurally similar, electromagnetic field) cannot be shielded,
thus every material object in the universe is subject to the influence of the gravitational field.\footnote{\citet[p.\ 56]{zeh2007} discusses an interesting thought experiment originally due to Borel that shows how the microstate of a gas in a vessel under normal conditions on Earth will be completely changed, within seconds, as a result of the displacement of a mass of a few grams at around the distance of Sirius.} Further, entangled correlations between spatially separated systems as described by quantum theory also cannot be `shielded' and moreover do not decrease with distance \citep{herbst2015, yin2017}. Even in a near perfect vacuum like those which exist between galaxies, correctly describing a quantum system requires us to consider the vacuum fluctuations of the electromagnetic field \citep{zeh1970}. Thus, if we are to take seriously the idea that closed systems are fundamental, it seems that we should conclude, with Jonathan \citet[]{schaffer2013}, that ``the cosmos is the one and only fundamental thing'' \citeyearpar[p. 67]{schaffer2013}.\footnote{See also \citet[pp.\ 234, 243--244]{primas1990}.}

Of course, physics is not only concerned with describing the cosmos. But when the influence of the rest of the cosmos, i.e., of its environment, on a particular system of interest, $\mathcal{S}$, is negligible in the context of a particular investigation, then it is legitimate, for the purposes of that investigation, to treat $\mathcal{S}$ as closed \citep[see, e.g.,][p. 242]{wallaceIsolatedSystemsI}. Further, in the case where the influence of the environment is not negligible we can model the system fundamentally in terms of a dynamical coupling between $\mathcal{S}$ and a separate system $\mathcal{E}$ such that $\mathcal{S}$ and $\mathcal{E}$ together form a single closed system. In this sense, the \emph{closed systems view} has been highly successful in physics; and in theoretical frameworks formulated from the closed systems view such as Hamiltonian mechanics, results such as Noether's Theorem intimately relate the concept of a closed system to the existence of certain symmetries and corresponding conservation laws. The same can be said of classical electrodynamics \citep[though see][]{frisch2005}, and similarly for quantum theory in its standard formulation.

It is worth remarking, however, that the idea that the cosmos actually is a closed system has never been unanimously held. Isaac Newton, for instance, considered the possibility of (and even the need for) divine influence in the cosmos as a guard against it descending into chaos \citep[qu.\ 31, p.\ 402]{newton1730}. It is also worth remarking that the methodology of the closed systems view is of far less importance outside of the context of physics. Biology, notably, genuinely conceives of the systems it deals with as open, i.e., as under the active influence of their environment, and the same goes for large parts of economics,\footnote{Neoclassical economics, which represents markets as closed systems aiming at a state of equilibrium, is an exception here. See, e.g., \cite{Jakimowicz2020}.} political science, psychology, sociology and many other special sciences.\footnote{It has long been noted that living organisms require the exchange of certain substances with their environment to sustain their functions \citep[see, e.g.,][]{vonBertalanffy1950}. This led to the development of general systems theory in the 1960s \citep{vonBertalanffy1988}. Roughly at the same time, a number of related fields that similarly construe systems as genuinely open were established, including cybernetics \citep{wiener1950}, information theory \citep{shannon1949} and the theory of complex (adaptive) systems (see, e.g., \citet{ladyman2020, thurner2018} and \citet{waldrop1993}). Other theories in which open systems figure centrally are as diverse as Maturana and Varela's theory of autopoietic systems \citeyearpar{maturana1980}, Luhmann's systems theory \citeyearpar{luhmann2012} and Haken's synergetics \citeyearpar{haken1983} which has found many applications in the natural and social sciences. The notion of noise, a typical open systems phenomenon, also plays an important role in psychology and cognitive science \citep[see, e.g.,][]{kahneman2021}. Other examples include network theory \citep[see, e.g.,][]{jackson2010}, dynamical causal modeling \citep[]{weinberger2020}, complex systems theory \citep[see, e.g.,][]{thurner2018}, and the construal of agents as open systems in artificial intelligence (see, e.g., \citet{briegel2012}, \citet{dunjko2018} and \citet{russell2021}).
} 
One can say that unless there are special reasons for describing the entities of a particular scientific domain as closed, in general it makes more sense to conceive of them as open. Moreover there is no fundamental methodological requirement in these sciences to derive the dynamics of an open system from those of a larger closed system. Far from being paradigmatic of the practice of science, physics, conducted in accordance with the closed systems view, stands out as the lonely exception, or so it would seem.

The foregoing considerations motivate us to take a closer look at the status of the closed systems view within physics, both at the methodology according to which the dynamics of an open system $\mathcal{S}$ is fundamentally conceived in terms of a coupling between it and a separate system $\mathcal{E}$, as well as at the metaphysical conception of the cosmos as itself a closed system. Upon doing this we find that taking the closed systems view in physics is not without its \emph{prima facie} problems.
Although standard models of cosmology (based on the Friedman-Lema\^{i}tre-Robertson-Walker solutions to the Einstein field equation) do describe the universe as closed, they are in many cases based on strong idealizations introduced for little reason other than to simplify their associated mathematics \citep[Sec. 1.1]{smeenkEllisCosmologySEP}. Particularly noteworthy is the scale factor which, as \citet[]{sloan2021} points out, carries no physical significance in itself. Thus, although our best cosmological models describe our universe as a closed system, this does not necessarily mean that our universe actually is a closed system \citep[see also][]{grybSloanScaleSurplus, sloan2018}. As for Hawking's proposal to model a black hole's dynamics as formally similar to that of an open system \citep[]{hawking1976b}, this is, to be sure, controversial (for discussion, see \citealt{giddings2013, pageUniverseOpenSystem, wallace2020a}). But although our goal here is not to defend Hawking's particular model \emph{per se}, we think it is important to point out that at least part of the motivation for wanting to reject it seems to amount to nothing more than that it runs counter to quantum theory and the fundamental dynamics described by it (see, e.g., \citealt[pp.\ 32--34]{giddings2013}; cf. \citealt[p. 219]{wallace2020a}).\footnote{We will discuss this in more detail in Section \ref{s:phil-phys-impl}.}

It seems, then, that in order to get to the core of the matter \emph{vis \'a vis} the status of the closed systems view in physics we must turn our attention to quantum theory, which will accordingly be our principal concern in the rest of this paper. As we will see, there are actually two theoretical frameworks through which one can describe quantum phenomena. The first of these, formulated in accordance with the closed systems view, is what we will call \emph{standard quantum theory} (\textbf{ST}). The second, formulated in accordance with the \emph{open systems view}, is what we will call the \emph{general quantum theory of open systems} (\textbf{GT}). On the open systems view, open systems, i.e., systems evolving under the influence of their environment, are conceived of as fundamental, and the dynamics of an open system is not described in terms of a coupling between two systems; instead the influence of the environment on a system is represented in the dynamical equations that govern its evolution. Over the course of this paper we will argue that \textbf{GT} is a more fundamental theoretical framework than \textbf{ST}; that, just as it is in the special sciences, the open systems view is therefore more fundamental than the closed systems view in physics; and finally, that since there is no third option, the open systems view is fundamental \emph{tout court}, and that this has important implications for the foundations and philosophy of physics, the philosophy of science, and for metaphysics.

We will begin, in Section \ref{s:views-frameworks-qt}, by clarifying what we mean by the terms `theoretical framework,' `theory,' and `model,' and also what we mean by a `view'. Then in Sections \ref{s:st} and \ref{s:gt}, we present the conceptual cores of \textbf{ST} and \textbf{GT}, respectively. We discuss the main assumptions involved in the derivation of the Lindblad equation governing the evolution of an important class of open systems, for the case of \textbf{ST}, in Section \ref{s:lindblad-st}, and for the case of \textbf{GT} in Section \ref{s:lindblad-gt}. In Section \ref{s:wider-framework} we examine one of these (the complete positivity principle) in more detail, arguing that while it is reasonable to enforce it in a framework formulated in accordance with the closed systems view like \textbf{ST}, it is more natural, in general, not to enforce it in \textbf{GT}, thus illustrating that \textbf{GT} has more expressive power than \textbf{ST}. In Section \ref{s:interpreting-st-and-gt} we consider how to interpret \textbf{GT} and \textbf{ST}. In the case of \textbf{ST} we focus on interpretations that take it to be a candidate fundamental theoretical framework for physics (though not necessarily in the same sense). These are Everettian and orthodox interpretations, broadly construed, where the former include (but are not exhausted by) many-worlds interpretations, and the latter include (neo-)Bohrian, (neo-)Copenhagen, pragmatist, relational, and QBist interpretations. We argue that for both Everettian and orthodox interpretations, the subject matter of \textbf{ST}, like the subject matter of \textbf{GT}, includes open systems, despite the fact that \textbf{ST} is formulated in accordance with the closed systems view.

In Section \ref{s:ontic-fund} we explicate the concept of an object of a theoretical framework as that through which a framework's subject matter is represented, and we consider the relation of what we call relative ontic fundamentality as it obtains between frameworks. We argue, in turn, from the point of view of orthodox, Everettian, and hidden-variable interpretations (which reject \textbf{ST} as a candidate fundamental framework) that in each case one should conclude that \textbf{GT} is, ontologically speaking, a more fundamental framework than \textbf{ST}, along the way clarifying how the various aspects of the concept of a view work together.

Having argued that \textbf{GT} is more fundamental than \textbf{ST} if one cashes out the notion of relative fundamentality in ontic terms, we then consider, in Section \ref{s:alternative-funds}, the possible objection that the empirical success of \textbf{ST} undercuts the motivation to look for a more fundamental theoretical framework in the first place. We argue that, on the contrary, the empirical success of \textbf{ST} gives as much \emph{prima facie} support to \textbf{GT} as it does to \textbf{ST}. The question of empirical support is arguably not the only epistemic question relevant here, however. Further, there is the question of which of the two frameworks provides more fundamental explanations of phenomena. Motivated by this we then consider two alternative notions of fundamentality: epistemic fundamentality (Section \ref{s:epistemic-fund}) and explanatory fundamentality (Section \ref{s:explanatory-fund}). In the case of epistemic fundamentality we consider two possible explications of the concept: in terms of the relation of justification, and in terms of the order in which we actually come to know an object of a theoretical framework in the context of a particular empirical investigation. We argue that the latter explication is more appropriate in the context of comparing theoretical frameworks, and conclude on this basis that \textbf{GT} is epistemically more fundamental than \textbf{ST}. In the case of explanatory fundamentality we consider three alternative explications, none of which, we argue, is entirely satisfactory. We conclude, in this case, that {\em if} one wants to entertain the notion of explanatory fundamentality at all, then it is best to employ a deflationary explication; and since \textbf{GT} has been determined to be both ontologically and epistemically more fundamental than \textbf{ST}, it should (therefore) be considered to be explanatorily more fundamental than \textbf{ST} as well.

Since \textbf{GT}, formulated in accordance with the open systems view, is more fundamental than \textbf{ST}, formulated in accordance with the closed systems view, we conclude that the open systems view is more fundamental than the closed systems view in quantum theory, and in Section \ref{s:tout-court} we argue that we should consider it to be fundamental \emph{tout court}. We remark that our argument for the fundamentality of the open systems view is \emph{not reductive}. We do not, that is, conclude that the open systems view is fundamental simply because it is fundamental in quantum theory (or in physics more generally). Rather, our argument (as we explained above) is motivated in the first place by considering the special sciences. This leads us to investigate physics, what one might accordingly think of as the last refuge of the closed systems view, and to conclude, upon closer consideration, that it is nothing of the sort.

Having argued for the fundamentality of the open systems view we then, in Section \ref{s:implications}, discuss some of its wider implications. In Section \ref{s:phil-phys-impl}, we discuss other issues in physics and philosophy of physics that can be illuminated by taking the open systems view. We review the issues of the `arrow of time' and the `information loss paradox,' arguing that these can be usefully informed by taking the open systems view. In Section \ref{s:phil-sci-impl} we turn our attention to more general issues in the philosophy of science. We compare the concept of a view with the related concept of a stance \citep[]{vanFraassen2002}. We clarify how, unlike a stance, one can rationally argue for a view. Finally, in Section \ref{s:phil-impl} we discuss some of the wider implications of the open systems view for metaphysics, including what it implies for our conception of the cosmos as a whole. We find that, when it comes to the cosmos as a whole, the metaphysical concepts of open and closed systems would appear to break down, but they nevertheless do so in a way consistent both with the methodological and metaphysical presuppositions of the open systems view, and point the way forward to future (meta)physics.

\section{Views, Frameworks and Quantum Theory}
\label{s:views-frameworks-qt}

In the philosophical literature on scientific theories one regularly encounters three terms which are too often not carefully distinguished from one another: `framework,' `theory,' and `model,' here ordered according to their generality, i.e., with regard to how `far away' each is, conceptually, from the specific objects under consideration in a given field of inquiry (i.e., from `target systems'). Closest to the target system one finds the (theoretical) model, the main representational device used in science. A {\em model} represents a specific system for a particular epistemic or practical purpose such as explanation, prediction or the planning of an intervention. The model of the hydrogen atom, for instance, represents it in terms of a positively charged proton (=the atomic nucleus) with a certain mass $m_p$ and a negatively charged electron with a certain mass $m_e$.

Models are usually formulated in the context of a specific {\em scientific theory}.\footnote{Exceptions are {\em phenomenological models} \citep[see, e.g.,][]{friggSEP}, which account for a target system using a wide range of assumptions which do not necessarily arise from the same theory. Such models are an important part of science, and often pave the way to new theories \citep{hartmann1995, hartmann1999, kao2019, kaoForthcoming}, but for our purposes we need only be concerned with models formulated in the context of a single theory.} The model of the hydrogen atom, for instance, is typically formulated in non-relativistic quantum theory which allows us to calculate the different discrete energy levels of the atom. One could also use an alternative theory such as relativistic quantum theory, which would yield various relativistic corrections to the non-relativistic account.

We note that non-relativistic and relativistic quantum theory have much in common. In both, for instance, the state of a system is represented by a state vector and the time evolution of a state vector is unitary. This is because both are formulated within the same \emph{theoretical framework}: \textbf{ST}, which accounts for what both theories always assume, i.e., in all of their models.\footnote{\citet[p.\ 2]{nielsenChuang2000} draw the analogy between a framework and the particular theories it subsumes, on the one hand, and an operating system on which particular computer applications are run, on the other. We note that this usage of the term `framework' is close to Rudolf Carnap's. As \citet[sec.\ 1.2]{leitgebCarusSEP} explain: ``\dots a Carnapian framework is meant to reconstruct the conceptual and inferential presuppositions of a scientific theory rather than a scientific theory itself.'' It is also related to a Kuhnian paradigm; \citet{friedman2001}, for instance, has argued that a (Carnapian) framework is a successful rational reconstruction of a Kuhnian paradigm (see also \citealt{irzik1995}).} As we will explain in more detail in the next two subsections, \textbf{ST} is not the only theoretical framework for describing the physics of quantum systems. There is also the alternative framework, \textbf{GT}. As we will see, \textbf{ST} and \textbf{GT} are interestingly different, both metaphysically and methodologically. In particular, each framework is formulated from what we will be calling its corresponding {\em view}. Associated with a view are: (i)~a set of methodological presuppositions in accordance with which we characterize the objects of a given domain, that is (ii)~motivated by a particular metaphysical position concerning the nature of those objects.\footnote{Note that the same view might be applicable to more than one domain, e.g., physics and biology, in which case one will be able to formulate a theoretical framework corresponding to that view in both sciences.}

On the one hand, \textbf{ST} is formulated in accordance with the \emph{closed systems view}. This is the view motivated by the metaphysical position according to which isolated systems are thought of (\emph{a priori}) as exclusively making up the subject matter of a given scientific domain, and it is associated with the methodology that models the dynamics of a given open system, $\mathcal{S}$, in terms of its being coupled to a further system $\mathcal{E}$ (representing its environment), such that $\mathcal{S+E}$ form an isolated system. \textbf{GT}, on the other hand, is formulated in accordance with the \emph{open systems view}. This view is motivated by the metaphysical position according to which the subject matter of a given scientific domain is thought of (\emph{a priori}) to consist of systems that are in general open, i.e., in interaction with their environment.\footnote{Our concern in this paper is with scientific theories and theoretical frameworks but in principle a given view might have an even wider application. It might be interesting to cash out the difference between the open and closed systems views in the context of moral or ethical theories, for instance.}
Its methodology is such that, rather than modeling the dynamics of a given open system $\mathcal{S}$ in terms of an interaction between two systems, we instead represent the influence of the environment on $\mathcal{S}$ in the dynamical equations that we take to govern its evolution from one moment to the next.

In contrast to other related concepts that have been discussed in the philosophy of science literature,\footnote{We have in mind, in particular, van Fraassen's concept of a `stance' \citep[]{vanFraassen2002}, which we will return to in Section \ref{s:phil-sci-impl}.} we submit that one can argue for a particular view and rationally decide between competing views. Giving such an argument, in favor of the open systems view, will be the goal of Section \ref{s:fundamentality}. In the remainder of this section our goal will be to present the main conceptual features of \textbf{ST} and \textbf{GT}, respectively, and to relate what these frameworks purport to tell us about the world.

Before we move on we should stress that distinguishing between theories and frameworks is not always a straightforward exercise. It is, for example, well known that there are different versions of classical theory (associated with the names of Newton, Hamilton and Lagrange) and also of quantum theory (including the path integral `formulation'), and the question arises of how, in general, to determine whether two given formalizations of a given field of inquiry represent two different theories within the same theoretical framework, or two different frameworks entirely. There is much to say about this and other related questions,\footnote{For example it might be argued that the various theories formulated from within a given theoretical framework can be grouped into more or less distinct classes (relativistic quantum theories might be thought of in this way, for instance).} as well as about theoretical equivalence more generally, but for our purposes we can leave such general questions aside. Instead we will focus on a specific case for which, as we will see in more detail in the remainder of this section, the two frameworks under consideration (\textbf{GT} and \textbf{ST}) are distinct.

\subsection{Standard Quantum Theory (\textbf{ST})}
\label{s:st}

\subsubsection{The Framework}
\label{s:st-framework}

We begin with \textbf{ST}, the form of quantum theory presented in most textbooks on the topic. In \textbf{ST}, the physical state of a system, ${\cal S}$, is represented by a normalized \emph{state vector} $\vert \psi (t)\rangle$ (from here on any mention of `state vector' should always be understood to mean a \emph{normalized} state vector \citep[sec.\ 2.1]{ismaelSEP}, whether or not we explicitly make this qualification). The possible states of ${\cal S}$ are called its \emph{state space}, represented as an abstract Hilbert space \citep[sec.\ 2.2]{ismaelSEP}. $\mathcal{S}$'s dynamical evolution is determined by a Hamiltonian $H$: a linear operator whose eigenvalues are the possible energy values of the system. The Hamiltonian is assumed to be Hermitian (or self-adjoint) which guarantees that its eigenvalues are real \citep[sec.\ 2.2]{ismaelSEP}. In non-relativistic quantum theory the time evolution of a state vector $\vert \psi (t)\rangle$, acted on by a Hamiltonian $H$, is given by the {\em Schr\"odinger equation}:\footnote{We follow the convention to set $\hbar =1$.} 
\beq
\label{eqn:schroedinger}
{\rm i} \, \frac{\partial}{\partial t} \, \vert \psi (t)\rangle = H \, \vert \psi (t) \rangle .
\eeq

\textbf{ST} has many applications, which can be grouped according to what they have in common. For example, it is applicable to atoms (`atomic physics'), atomic nuclei (`nuclear physics'), and solid state systems (`condensed matter physics'), each of which is characterized by a certain class of target systems, Hamiltonians and modeling strategies. For example, within atomic physics, constructing a model of an atom involves specifying a Hamiltonian which is the sum of the Hamiltonians of the various subsystems and of the terms that account for their interaction. In the case of the hydrogen atom, for instance, the Hamiltonian consists of three terms: a kinetic energy term representing the nucleus (i.e. the proton), a kinetic energy term representing the electron, and a term accounting for the Coulomb attraction of the electron and the proton. More sophisticated models are also possible, taking, for instance, the interaction between the magnetic field from the electron movement and the nuclear spin into account. Once the Hamiltonian is specified, one can solve the corresponding Schr\"odinger equation.
Note that a system such as the hydrogen atom can also be analyzed using relativistic quantum theory. In this case, one solves the Dirac equation instead of the Schr\"odinger equation.

\textbf{ST} has a number of important features that apply to every model of every theory formulated within it. Here are four of them:
\benum
\item \emph{The time evolution of a state is unitary}. In nonrelativistic quantum theory (one can give a parallel argument for relativistic quantum theory) this follows from the fact that the formal solution of Eq.\ (\ref{eqn:schroedinger}) is $\vert \psi (t) \rangle = U(t) \, \vert \psi (0)\rangle$ with the time evolution operator $U(t) = \exp (- {\rm i} \, H \, t)$. From this equation and the Hermiticity of $H$ it follows that $U(t)$ is unitary, i.e. $U(t)$ satisfies $U(t) \, U(t)^{\dag} = U(t)^{\dag} \, U(t) = I$ where $U(t)^{\dag}$ is the adjoint of $U(t)$ and $I$ represents the identity operator (note that this implies that unitary operations are reversible).
\item \emph{All phenomena are modeled in terms of closed systems}. This follows from the fact that a target system's Hamiltonian includes no terms reflecting its interaction with an external environment. It also does not include non-unitary terms which would represent the loss of a conserved quantity. In particular:
\item \emph{Energy is conserved} in the sense that the expectation value $E = \langle \psi \vert H \vert \psi \rangle $ is a constant of motion. This follows directly from the unitarity of the time evolution of the state vector.\footnote{For further discussion, see \citet[sec.\ 3.1]{maudlinetal2020}.}
\item Finally, {\em probability is conserved}. To see why, note that the probability $p_m(t)$ of obtaining the outcome $m$ given a measurement at time $t$ on a system in the state $\vert \psi (t) \rangle$ is:
  $$p_m(t) = \langle \psi(t) | M_m^\dagger M_m  | \psi(t) \rangle,$$
  where $M_m$ is the measurement operator corresponding to $m$ (note that for a projective measurement,\footnote{A projective measurement can be thought of as an ideal or perfect measurement and corresponds to the case for which the various outcomes are perfectly distinguishable from one another \citep[see][sec.\ 2.4]{hughes1989}. We will discuss the issue of measurement in more detail in Section \ref{s:interpreting-st-and-gt}.} $M^\dagger_m M_m = M_mM_m = M_m$), and the measurement operators corresponding to the various outcomes satisfy the completeness relation $\sum_m M^\dagger_m M_m = I$ \citep[see also][p.\ 85]{nielsenChuang2000} which expresses that the various probabilities sum to one.\footnote{$\langle \psi(t) | I | \psi(t) \rangle$ = $\langle \psi(t) | \psi(t) \rangle$, and the inner product of a normalized state vector with itself is always 1.} Probability conservation follows from unitarity; in particular:
  \begin{align}
    \label{eqn:prob-conserv}
    \textstyle\sum_m p_m(t) =\; & \textstyle\sum_m \langle \psi(t) | M^\dagger_m M_m | \psi(t) \rangle = \langle \psi(t) | I | \psi(t) \rangle \nonumber  \\
    =\; & \langle \psi(0) | U^{\dag}(t) \, U(t) | \psi(0) \rangle = \langle \psi(0) | I | \psi(0) \rangle \nonumber \\
    =\; & \textstyle\sum_m \langle \psi(0) | M^\dagger_m M_m | \psi(0) \rangle \nonumber  \\
    =\; & \textstyle\sum_m p_m(0) = 1,
  \end{align}
  though we note that \emph{unitarity is only a sufficient condition} for the conservation of probability. We will see in Section \ref{s:lindblad-st} that probability is also conserved if the system's dynamics are non-unitary and of a certain form. 
\eenum

Systems as described by \textbf{ST} may be as big as one likes (in principle a system can include everything that exists),\footnote{This assumes that one thinks it makes sense to model such a system. \textbf{ST} is agnostic on this issue in the sense that its role as a conceptual framework is to provide the conceptual tools with which to construct a dynamical model of whatever we take to be a system of interest. It is up to particular theories to then take up the question of how, using the language of \textbf{ST}, to model particular classes of systems, assuming it makes sense, according to the theory, to model them at all.} and will typically be composed of more than one subsystem. In some cases such a system will be in a \emph{product state}, which means that each of its subsystems, in addition to the system as a whole, is assigned its own state vector. For instance, the general form for the product state of a system made up of two two-dimensional subsystems $\mathcal{A}$ and $\mathcal{B}$ is given by:
\begin{align}
  \label{eqn:product-state}
  | \Psi \rangle_{\mathcal{A+B}} \;=\; | \psi \rangle_{\mathcal{A}} \,\otimes\, | \phi \rangle_{\mathcal{B}},
\end{align}
where $| \psi \rangle_{\mathcal{A}}$ and $| \phi \rangle_{\mathcal{B}}$ are normalized state vectors assigned to $\mathcal{A}$ and $\mathcal{B}$, and $\otimes$ is the tensor product symbol \citep[see][sec.\ 2.1]{ismaelSEP}.\footnote{We follow the convention to use capital Greek letters when the state of a combined system is being referred to and lowercase Greek letters otherwise.} States of this form comprise a relatively small portion of the combined state space, $\mathcal{H}_{\mathcal{A}} \otimes \mathcal{H}_{\mathcal{B}}$, of $\mathcal{A}$ and $\mathcal{B}$. In general, a combined system will be in an \emph{entangled state}. For instance, the state
\begin{align}
| \Psi^{\,\mbox{-}} \rangle_{\mathcal{A+B}} \;=\; \frac{1}{\sqrt 2}\big(| \psi \rangle_{\mathcal{A}} \,\otimes\, | \phi \rangle_{\mathcal{B}} \;-\; | \phi \rangle_{\mathcal{A}} \,\otimes\, | \psi \rangle_{\mathcal{B}}\big)
\end{align}
cannot be rewritten in the form given by Eq.\ \eqref{eqn:product-state}, thus there is no way to assign normalized state vectors to its individual subsystems $\mathcal{A}$ and $\mathcal{B}$.

It is sometimes said that for a system in a state like $| \Psi^{\,\mbox{-}} \rangle_{\mathcal{A+B}}$, the whole is prior to its parts \citep[see, e.g.,][]{howard1989, schaffer2010, schaffer2013}, which is another way of saying that the properties exhibited by $| \Psi^{\,\mbox{-}} \rangle_{\mathcal{A+B}}$ are not truly describable in terms of parts at all.
Given, that is, that there is no way to assign normalized state vectors to the individual subsystems of the system, it seems we must either maintain that they cannot be fully described within \textbf{ST}---in which case \textbf{ST} is incomplete---or else we must maintain that the only system being described in a case like this is the entangled system as a whole. Both options amount to the same thing with regard to the concept of a subsystem of an entangled system: Since it cannot be assigned a state vector, there would strictly speaking seem to be no place for this concept within \textbf{ST} whichever of the above two options we are inclined toward.\footnote{Einstein's argument that quantum theory is incomplete relies on similar logic \citep[]{ramirez2020}. It follows from Einstein's incompleteness argument that we must accept either (i)~that measurements in one region of space may change `telepathically' the real state of a physical system in some distant region of space, or (ii)~that distant regions of space cannot in general be assigned real states. Einstein rejects both alternatives and concludes that quantum theory is an incomplete theory. Taking the real dynamical state of a closed quantum system to be represented by a unitarily evolving state vector requires that we accept (ii), since there is no further dynamical law that one can then appeal to that can be responsible for (i). To get (i), one must posit some sort of collapse mechanism, either in the form of a bare posit (i.e., as in an ontologically literal interpretation of the projection postulate) or in the form of a further dynamical mechanism \citep[]{ghirardi2018}. Either way this amounts to a more general evolution of the state vector than unitary evolution. Compare also \citet[sec.\ 5]{primas1990}.} Appearances are misleading, however, as we will see in more detail in Section \ref{s:interpreting-st-and-gt}.

\textbf{ST} is a highly successful theoretical framework. It can be used to study a large range of systems, including the atoms, nuclei, and condensed matter systems we mentioned above. For these, the closed systems view seems physically plausible insofar as it has been experimentally confirmed that such systems can be effectively isolated at least to a large degree, and that factoring in the influence of their environment yields in most cases only a negligible contribution to whatever quantities are of interest. Other contributions, such as the effects of the interaction of the electron magnetic field with the nuclear spin in the case of the hydrogen atom, are much more important, but are still describable in terms of closed systems within \textbf{ST} in a natural way. Not all phenomena can be modeled in a simple way as closed systems, however. Some phenomena are decidedly `open systems phenomena' such that the dynamics of a given target system needs to be modeled as effectively determined to a considerable extent by the system's environment. Lasers are an example: A laser is pumped by an external energy source, and coherent laser radiation is in turn coupled out of the system. The spontaneous emission of a photon by an atom is another. We turn to the question of how such processes can be described within \textbf{ST} in the next subsection.

\subsubsection{Open Systems}
\label{s:lindblad-st}

An important historical predecessor of the quantum theory of open systems (as formulated within \textbf{ST}) is the Weisskopf-Wigner theory of spontaneous emissions \citep{weisskopfwigner1930}, which accounts for the observation that atoms which have been prepared in an excited state will randomly emit photons over time, returning to their ground state while the emitted radiation disappears into the universe. The starting point of the theory is that while the exact time of emission is impossible to determine, it is experimentally well known that an excited state decays exponentially at a certain rate. Given this, questions such as how the decay rate depends on particular characteristics of the atoms, which characteristics are relevant, and what the decay mechanism is, arise. But giving a closed systems account of spontaneous emission is problematic since the radiation is sent into empty space and there are infinitely many modes and directions by which this can occur. Instead, in the Weisskopf-Wigner theory one models the decay as a stochastic process. Although Weisskopf and Wigner did not explicitly refer to this as a theory of open systems, theirs was the first contribution to what has since been further developed into the powerful, elegant, and more general theory which we now introduce. The basic strategy employed is to embed the open system of interest within a larger system, i.e., a sufficiently large `box' that guarantees that no more than a negligible amount of a given quantity of interest can flow out.

Before we say more let us first recall a few further basic quantum-mechanical concepts related to open systems that we will be making use of later. To begin with we consider a large number---usually called an \emph{ensemble}---of similar quantum systems $\mathcal{S}_i$ which have all been prepared using an identical \emph{preparation procedure}.\footnote{Although we are using the concept of a quantum statistical ensemble to flesh out the meaning of the density operator, we are not in any way committing ourselves to frequentist interpretation of probability \citep[cf.][p.\ 151]{myrvold2021}. The idea of an ensemble is only a particularly useful fiction for conveying the difference between pure, properly mixed, and improperly mixed states, and what it means for a properly and an improperly mixed state to be (locally) indistinguishable from one another.} For instance one might specify that a given system, $\mathcal{S}_i$, is to be prepared in the state $| \psi_1 \rangle$ with probability $p$ and in the state $| \psi_2 \rangle$ with probability \mbox{$1-p$}, which will progressively generate an ensemble for which the relative frequencies of $| \psi_1 \rangle$ and $| \psi_2 \rangle$ tend, respectively, toward $p$ and $1-p$ as we prepare more and more systems. The so-called \emph{properly mixed state} of such an ensemble is represented by a \emph{density operator} \citep[see also][ch.\ 5]{hughes1989}, which can be expressed as follows:
\beq
\rho = p \, | \psi_1 \rangle \langle \psi_1 | \;+\; (1-p) \, | \psi_2 \rangle \langle \psi_2 |.
\eeq

In addition to representing the state of the ensemble, we can also think of $\rho$ as characterizing each of its members (indeed, the very idea of an ensemble can be regarded as a useful fiction to aid us in characterizing an individual system in this more general way), since each member of the ensemble has been identically prepared. When there are more than two alternatives, $\rho$ is given more generally by
\beq \label{properly-mixed-state}
\rho = \sum_j p_j \, | \psi_j \rangle \langle \psi_j |,
\eeq
where the $p_j$ are non-negative real numbers summing to 1. Note, however, that the relation between preparation procedures and ensembles (and their corresponding $\rho$) is many-to-one: For a given ensemble whose state is represented by some density operator $\rho$, there is in general more than one preparation procedure that will give rise to it \citep[p. 103]{nielsenChuang2000}; i.e.,
\begin{align}
  \label{eqn:non-uniqueness-1}
  \rho & = \sum_j p_j \, | \psi_j \rangle \langle \psi_j \vert \\
  \label{eqn:non-uniqueness-2}
  & = \sum_k p'_k \, | \phi_k \rangle \langle \phi_k \vert.
\end{align}
Despite this, the predictions yielded by the density operator $\rho$ are the same regardless of how it is decomposed into state vectors; i.e., the right-hand sides of Eq.\ \eqref{eqn:non-uniqueness-1} and Eq.\ \eqref{eqn:non-uniqueness-2} predict the same statistics for measurements.

The one exception to the many-to-one rule is the case of an ensemble for which every system is prepared in the same state $| \phi \rangle$. In this case we characterize the ensemble using a \emph{pure state} whose density operator takes the form:
\beq \label{pure-state}
\rho = | \phi \rangle \langle \phi |.
\eeq

There are also more general ways to generate an ensemble. If one were to correlate pairs of quantum particles, for instance, and then ignore the right-hand particle from each pair, the ensemble of left-hand particles progressively generated in this way would be in what is called an \emph{improperly mixed state} \citep[]{despagnat1966, despagnat1971}. Because correlated quantum systems are generally entangled in quantum theory, an ensemble in an improperly mixed state, $\rho$, \emph{cannot} be understood as having been generated using a probabilistic procedure like the ones given above; i.e., a system in such an ensemble cannot be understood to be in a state $| \psi_j \rangle$ with a certain probability. This is even though the statistics arising from a sequence of measurements on the members of an improperly mixed ensemble are effectively indistinguishable from a sequence of those same measurements on a properly mixed ensemble whose state is also described by $\rho$.

This is the kind of case one deals with in the quantum theory of open systems, where the improperly mixed state characterizing the system of interest $\mathcal{S}$, represented by the \emph{reduced density operator} $\rho_{\mathcal{S}}$, is derived by tracing over the degrees of freedom of the environment from the combined state $| \Psi \rangle_{\mathcal{S+E}}$. This amounts to preparing an ensemble of systems $\mathcal{S+E}$, all in the state $| \Psi \rangle_{\mathcal{S+E}}$, and then selecting $\mathcal{S}$ from each pair (and ignoring $\mathcal{E}$) to form a new ensemble in an improperly mixed state represented by the reduced density operator $\rho_{\mathcal{S}}$.

Let us now see how the machinery just outlined can be applied to a specific target system.\footnote{For two modern textbooks on the topic, see \citet[]{alicki2007, breuer2007}. \citet[]{davies1976} is a classic text.} We consider a single two-level atom ($= {\cal S}$) that is coupled to an environment ${\cal E}$, such that ${\cal S + E}$ form a closed system. To account for its dynamics, we make the following assumptions:
\benum
\item The total Hamiltonian is given by $H = H_S + H_E + H_{SE}$. Here $H_S$ is the Hamiltonian of the system (i.e. the two-level atom), $H_E$ is the Hamiltonian of the environment which is modeled as an infinite collection of two-level atoms. Finally, $H_{SE}$ accounts for the interaction between the system and the environment. 
\item ${\cal S}$ and ${\cal E}$ are initially uncorrelated and weakly coupled, i.e. ${\cal E}$ affects the state of ${\cal S}$ but not vice versa. This is sometimes called the {\em Born approximation}.
\item The future state of ${\cal S}$ only depends on its present state. Metaphorically put, ${\cal S}$ has a short memory and `forgets' its earlier states. This is the {\em Markov approximation}.
\eenum

After carrying out the calculation, the variables representing the environment are then `traced out.' This yields an equation for the evolution of the reduced density operator $\rho$ in the quasi-spin formalism:
\beq \label{lindspec}
\dot{\rho} = -{\rm i} \, [H_S, \rho] + A \, \left( [\sigma_{-} \, \rho, \sigma_{+}] + [\sigma_{-}, \rho \, \sigma_{+}] \right).
\eeq
Here the basis states are $| 1\rangle := (1, 0)^{\rm T}$ and $| 0\rangle := (0, 1)^{\rm T}$ and the Pauli matrices $\sigma_{\pm}$ represent the corresponding raising and lowering operators.\footnote{That is, $\sigma_{+} | 0\rangle = | 1\rangle$, $\sigma_{+} | 1\rangle = 0$, $\sigma_{-} | 0\rangle = 0$ and $\sigma_{-} | 1\rangle = | 0\rangle$.} $A$ is the decay rate of the target system. 

It is interesting to note that the dynamics of $\rho$ in Eq.\ (\ref{lindspec}) involves a unitary part and a non-unitary part. The unitary part is given by ${\cal L}_{u} \, \rho := -{\rm i} [H_S, \rho]$ where we have introduced the superoperator ${\cal L}_{u}$. 
Abbreviating the non-unitary part by ${\cal L}_{n-u} \, \rho$, Eq.\ (\ref{lindspec}) can be written in the compact form,
\beq \label{lindrhoc}
\dot{\rho} = ({\cal L}_{u} + {\cal L}_{n-u}) \, \rho.
\eeq
In the absence of the non-unitary part, Eq.\ \eqref{lindrhoc} is equivalent to the Liouville-von Neumann equation which reduces to the Schr\"odinger equation (or its relativistic counterpart) if the initial state of the system is a pure state. Note that the specific form of the non-unitary part in Eq.\ \eqref{lindspec} depends on the particular system under consideration. If one were to consider a different system, e.g. a decaying photon mode, the non-unitary part would look different. It turns out, however, that it always has the general form
\beq \label{lindrhoc1}
{\cal L}_{n-u}\, \rho = \frac{1}{2}\, \sum_i \left( [L_i \, \rho, L_i^\dag] + [L_i , \rho \, L_i^\dag]    \right),
\eeq
where the $L_i$ are (bounded) operators. Eq.\ \eqref{lindrhoc}, with the form of ${\cal L}_{n-u}$ given by Eq. \eqref{lindrhoc1}, is known as the Lindblad equation. It can be shown that if the non-unitary dynamics of a system has the form specified in Eq.\ (\ref{lindrhoc1}), then probability is conserved.\footnote{\label{f:prob-conv}Proof: We take the trace on both sides of Eq.\ (\ref{lindrhoc}) and obtain ${\rm Tr} \, \dot{\rho} (t) =  {\rm Tr} \left({\cal L}_{u} \, \rho \right) + {\rm Tr} \left({\cal L}_{n-u} \, \rho \right)$. As both terms on the right hand side are (sums of) commutators (see Eq.\ (\ref{lindrhoc1})) and the trace of a commutator vanishes, we get that ${\rm Tr} \, \dot{\rho} (t) = 0$. This implies that $d/dt \left({\rm Tr} \, \rho (t) \right)= 0$ and therefore ${\rm Tr} \, \rho (t) = {\rm Tr} \, \rho (0) = 1$. Next, following the derivation of Eq. (\ref{eqn:prob-conserv}), we calculate $\textstyle\sum_m p_m (t) = \textstyle\sum_m Tr  \left(\rho (t) \, M^\dagger_m M_m \right) = Tr \left(\textstyle\sum_m \rho (t) \, M^\dagger_m M_m \right) = Tr  \, \rho (t) = {\rm Tr} \, \rho (0) = 1$. Hence, $\textstyle\sum_m p_m (t) = \textstyle\sum_m p_m (0) = 1$.\qed} Probability conservation therefore does not presuppose an underlying unitary dynamics, confirming what we said in Section \ref{s:st-framework}.

Let us take stock of where we stand so far. We have seen that there are a number of physical phenomena (such as spontaneous emission and lasers) that require us to model the dynamics of a target system as driven to a considerable degree by its environment. The key idea which allows one to account for such phenomena within \textbf{ST} is to embed the open system under consideration into a larger closed system that includes, as its subsystems, both the system and its environment. One can then derive equations of motion describing the effective dynamics of the system that are non-unitary in general. Interestingly, given the assumptions listed above, the non-unitary part of the dynamics reflected in these equations always has the same form---the Lindblad form given in Eq.\ \eqref{lindrhoc1}---%
which suggests that a more principled discussion of open quantum systems is in order. This is the subject of the next subsection.

\subsection{The General Quantum Theory of Open Systems (\textbf{GT})}
\label{s:gt}

In the last subsection we outlined how to derive an equation of the Lindblad form for the dynamics of the specific open system that we were considering: a single two-level atom $\mathcal{S}$ evolving under the influence of its environment. We saw that one begins by considering the dynamics of the larger closed system $\mathcal{S+E}$ under certain constraints on how $\mathcal{S}$ and $\mathcal{E}$ interact with one another that can be reasonably assumed to hold. After evolving $\mathcal{S+E}$ forward, one then traces over the degrees of freedom of $\mathcal{E}$ to yield an equation for the reduced dynamics of $\mathcal{S}$. It turns out that following this procedure is not the only way to show that the dynamics of $\mathcal{S}$ are governed by an equation of the Lindblad form. This form can also be derived, as we will now see, by imposing certain `principles' or conditions on the dynamics of an open quantum system as generally construed. But while the general and specific derivations all agree with regard to the form of the equation governing the dynamics of a given open system $\mathcal{S}$, the assumptions involved in the general derivation of the Lindblad equation represent restrictions on the possible dynamics of systems as described within a theoretical framework that is distinct from \textbf{ST}. This framework is \textbf{GT}.

In \textbf{GT}, the physical state of a system at a given moment in time is not represented by a state vector. It is represented by a density operator that evolves, in general, non-unitarily over time. Unlike \textbf{ST}, for which the dynamics of an open system of interest, ${\cal S}$, is obtained via a contraction (through the partial trace operation discussed in the last subsection) of the total dynamics of $\mathcal{S+E}$ to the state space of $\mathcal{S}$; in \textbf{GT}, the dynamical equations that govern the evolution of ${\cal S}$ pertain directly to $\mathcal{S}$ itself. Systems are modeled as genuinely open in \textbf{GT}; i.e., \textbf{GT} is a framework formulated in accordance with the open systems view, according to which we do not describe the influence of the environment on $\mathcal{S}$ in terms of an interaction between two systems, but instead represent the environment's influence in the dynamical equations that we take to govern the evolution of $\mathcal{S}$.\footnote{The earliest work on characterizing the general form of the dynamics of a quantum system represented by a density operator---what we are here calling \textbf{GT}---is to be found in \citet{sudarshan1961} and in \citet{jordan1961}, who characterize the possible dynamical evolutions of a density operator as most generally given in the form of a linear map. Note that we do not want to claim that these physicists explicitly appeal to what we are here calling the open systems view or the metaphysical position that motivates it (see Section \ref{s:views-frameworks-qt}), though see \citet{weinberg2014} for an explicit recommendation to do away with the state vector as a representation of the state of a quantum system.}

\subsubsection{Deriving the Lindblad Equation in \textbf{GT}}
\label{s:lindblad-gt}

We now motivate and explain the main assumptions needed to derive the Lindblad equation for the evolution of an important class of open systems in \textbf{GT}.\footnote{We are focusing on the Lindblad equation in this section as it is particularly useful for illustrating the role that various physical principles like, for instance, complete positivity, play in the derivation of the dynamics of an open system. We are by no means claiming, however, that the Lindblad equation provides the most general description of open systems dynamics possible in \textbf{GT}. Most completely positive trace-preserving maps on the state space of an open system will, in fact, not be of this form \citep[]{wolfCirac2008}.} To begin with, we consider the evolution of a system $\mathcal{S}$ during a given period of interest as governed by a family of dynamical maps, such that a given map in the family, $\Lambda_t$, maps $\mathcal{S}$ from some initial state $\rho_{t_i}$ to $\rho_{t_f}$ where $t_f = t_i + t$. For a particular $\Lambda_t$ to serve its purpose as a dynamical map, it is reasonable to require that the state of $\mathcal{S}$ at $t_f$ be completely determined given only a full specification of both $\Lambda_t$ and of the system's state at time $t_i$. This is no different from, for instance, the situation in classical mechanics, where the state $\omega_{t_1}$ of a system at time $t_1$ can be fully determined through applying the classical dynamical laws to the full specification of its state, $\omega_{t_0}$, at time $t_0$. In the case of classical mechanics, a system's dynamics are reversible; i.e., the dynamical laws of classical mechanics will take the inverse of the system's state at time $t_1$, $-\omega_{t_1}$, into the state $\omega_{t_0}$ at time $t_2$.

State vectors in the quantum-theoretical framework represented by \textbf{ST} are, unlike state descriptions in classical mechanics, irreducibly probabilistic in the sense that they encode, in general, only the probability that a system will be found to have a particular value of a specified (observable) parameter given an ascertainment of that parameter. But with respect to these conditional probability assignments, \textbf{ST} is just like classical mechanics insofar as the evolution of the state of a quantum system in \textbf{ST} is deterministic; the state vector $| \psi(t_1) \rangle$ of a system at $t_1$ is determined through applying (in the case of non-relativistic quantum theory) the Schr\"odinger equation to the state vector $| \psi(t_0) \rangle$ of the system at $t_0$. And since the dynamics of systems in \textbf{ST} are unitary they are, just as in classical mechanics, reversible.\footnote{For a discussion of some of the conceptual subtleties associated with time reversibility in \textbf{ST} see \citet[]{robertsReversingBook}.}

In \textbf{GT}, in contrast, the dynamics of a system are non-unitary, and therefore irreversible, in general. Since reversibility is a feature of perfectly deterministic processes, the dynamics of systems in \textbf{GT} are not in general deterministic in that sense. If, however, every map $\Lambda_t$ associated with $\mathcal{S}$ is such as to uniquely map each state in its state space to another state in the same space, then the dynamical evolution of $\mathcal{S}$ thus described is an example of a \emph{Markov process}, a generalization of a deterministic process for which the probabilistic state of a given system at time $t_f$ is wholly determined, given $\Lambda_t$, by considering its probabilistic state at time $t_i$, though the reverse is not true in general. It follows from this assumption that the dynamical maps describing a system's dynamics may be mathematically composed. That is, describing the evolution of the state of a system during a period $t+s$ is equivalent to describing, first, its evolution during the period $s$, followed by its evolution during the period, $t$:
\begin{align}
  \label{eqn:markov}
  \Lambda_{t+s}\,\rho = \Lambda_t\Lambda_s\,\rho.
\end{align}

In order to complete the derivation of the Lindblad equation we require two further general assumptions. The first is that the evolution of an open system should at all times be `completely positive.' This is a subtle requirement, which we can both motivate and explain through the following argument. Any dynamical map, $\Lambda_t$, on the state space of a system should be such as to map one valid physical state of the system into another. Formally, a density operator is a \emph{positive semi-definite operator}, which effectively means that it assigns non-negative probabilities (as one should!) to the outcomes of measurements on a system. Further, since the probabilities of measurement outcomes must sum to 1, density operators must always be of unit trace: ${\rm Tr}(\rho) = 1$. It follows from this that $\Lambda_t$ must be a trace-preserving positive map; i.e., it must map one positive semi-definite operator of unit trace into another if it is to be physically meaningful.

In addition to describing the evolution of ${\cal S}$, it should also be possible to describe the evolution of ${\cal S}$ in the presence of further systems. Imagine, in particular, that ${\cal S}$ is evolving in the presence of its environment, ${\cal E}$, over a period of time, and that at the initial time, $t_0$, ${\cal S}$ and ${\cal E}$ are uncorrelated. Imagine, further, that there exists a `witness' system ${\cal W}_n$ (where $n$ is the dimensionality \citep[sec.\ 2.1]{ismaelSEP} of the witness's state space) that, let us assume, is inert and not evolving. We assume in addition that ${\cal S}$ and ${\cal W}_n$ are spatially separated and not interacting with one another presently. Then it suffices to describe the dynamics of ${\cal S}$ and ${\cal W}_n$, under these assumptions, if we trivially extend the dynamical map $\Lambda_t$ for ${\cal S}$ via the identity transformation $I_n$ on the state space of ${\cal W}_n$:
$$\Lambda_t \otimes I_n.$$
Requiring that $\Lambda_t$ be \emph{completely positive} means, not only that $\Lambda_t$ should be a positive map; but that, in addition, $\Lambda_t \otimes I_n$ should be a positive map for all $n$. Not all positive $\Lambda_t$ are completely positive in this sense, however.\footnote{One example of a positive but not completely positive map takes a density operator (expressed as a matrix) to its transpose \citep[see][p.\ 369]{nielsenChuang2000}.} Requiring complete positivity means that any such maps cannot be valid maps on ${\cal S}$. This is arguably a reasonable requirement on a dynamical map, for it seems reasonable to require that it should be possible to consider the action of $\Lambda_t$ on ${\cal S}$ regardless of ${\cal S}$'s initial state; in particular that the effect of $\Lambda_t$ on ${\cal S}$ should be the same, and should always be valid, regardless of the existence of some other system ${\cal W}_n$ with which $\mathcal{S}$ is not even interacting.

The last condition that we are going to require of our family of dynamical maps is also the simplest to state: In order to derive the Lindblad equation in the framework of \textbf{GT}, we require that it describe the system ${\cal S}$ as evolving continuously in time. Intuitively, this means that $\Lambda_t$ evolves $\rho_0$ so that, as $t$ approaches zero, the resulting state of the system becomes `infinitesimally close' to $\rho_0$.

A family of dynamical maps that satisfies all these requirements is called a \emph{quantum dynamical semigroup} (QDS).\footnote{A \emph{semigroup} is a generalization of the concept of a group that drops the requirements of identity and invertibility.} For a given QDS there exists a densely defined superoperator ${\cal L}$ which `generates' the semigroup in the sense that by successively applying ${\cal L}$ to a given density operator $\rho$ for a system at some initial time we can construct the dynamical map $\Lambda_t$ on $\rho$ for any $t$.\footnote{The analogue of the generator of a QDS in \textbf{ST} is the system's Hamiltonian, which likewise can be used to generate a (unitary) map on the state space of the system for any $t$.} One can show \citep[]{lindblad1976} that for a system $S$ characterized by a separable Hilbert space, the (bounded) generator ${\cal L}$ of a quantum dynamical semigroup for describing its motion can always be expressed in the general form given by Eqs.\ \eqref{lindrhoc}--\eqref{lindrhoc1}, which we derived for the specific case of a single two-level atom above.\footnote{The Lindblad equation is sometimes referred to as the Gorini--Kossakowski--Sudarshan--Lindblad (GKSL) equation, as \citet[]{gks1976} independently proved that Eq.\ \eqref{lindrhoc} is the general form of the generator for a system whose Hilbert space is finite-dimensional (though note that this is a special case of what was proven by Lindblad). Note that Lindblad's result does not hold for the case of an unbounded generator, but there are formally similar equations that can be used instead in all known cases \citep[][p.\ 8]{alicki2007}.}

The Lindblad equation gives the general form of the (bounded) generator of a continuous, completely positive, one-parameter dynamical semigroup describing the evolution of an open quantum system. Such evolutions are important and have many applications, including lasers and the spontaneous emission processes that we discussed, in the context of \textbf{ST}, in Section \ref{s:lindblad-st}.
There is more to \textbf{GT} than the Lindblad equation, however. In the next section we illustrate this by considering, as an example, the reasons for requiring one of the principles involved in its derivation---complete positivity---more closely. We will see that the reasons for requiring complete positivity are only compelling on the closed systems view, and that on the open systems view---the view from which \textbf{GT} is formulated---it is more natural not to impose this condition in general. Relaxing complete positivity thus results in possible dynamical descriptions of the evolution of a quantum system that have a natural interpretation in \textbf{GT} but which do not make physical sense in \textbf{ST}.

\subsubsection{A Wider Framework}
\label{s:wider-framework}

The earliest work on characterizing the general form of the dynamics of a quantum system represented by a density operator did not impose the requirement of complete positivity. Instead, \citet{sudarshan1961} and \citet{jordan1961} characterized the possible dynamical evolutions of a density operator as most generally given in the form of a linear map. Such maps are positive, but not necessarily completely positive. With the development of the theory of completely positive maps by \citet[]{choi1972}, \citet{gks1976}, \citet{lindblad1976}, \citet[]{kraus1983}, and others, however, a debate arose among physicists over whether to consider complete positivity as a fundamental dynamical principle. There is no need to review all of the details of this, at times passionate, debate here;\footnote{See, for instance, the exchange between \citet[]{simmons1981, simmons1982} and \citet[]{raggio1982}. A detailed account of the important exchange between \citet[]{pechukas1994, pechukas1995} and \citet[]{alicki1995} is given by \citet[][]{cuffaroMyrvold2013}, who also suggest a deflationary way to resolve the impasse. \citet{schmid2019} relate the debate over complete positivity to the literature on causal models \citep[for further discussion, see][]{evans2021}.} instead we limit ourselves to \citeauthor[]{shaji2005}'s \citeyearpar{shaji2005} criticism of what has become the standard argument for complete positivity: the `witness' argument that we reviewed in Section \ref{s:lindblad-gt}. Without repeating the details, recall that what makes the argument compelling is that the condition of complete positivity seems to follow from nothing more than a very commonsense requirement: that the validity of a dynamical map on the state space of a system, $\mathcal{S}$, should not depend on the non-existence of another (inert) system ${\cal W}_n$ that may in principle be very far away from $\mathcal{S}$ and not even interacting with it.

As \citeauthor[]{shaji2005} point out, however, this argument only works if $\mathcal{S}$ is actually entangled with $\mathcal{W}_n$.\footnote{Cf. \citet[p.\ 245]{primas1990}.} As long as $\mathcal{S}$ and $\mathcal{W}_n$ are not entangled, a positive but not completely positive map on $\mathcal{S}$'s state space will always yield a valid description of the evolution of $\mathcal{S}+\mathcal{W}_n$ under the remaining assumptions characterizing the setup (i.e., $\mathcal{W}_n$ is separated from $\mathcal{S}$, not interacting with $\mathcal{S}$, and evolving trivially). If, on the other hand, $\mathcal{S}$ and $\mathcal{W}_n$ are entangled with one another, then ``there must have been some sort of direct or indirect interaction between the two at some point and hence $\mathcal{W}_n$ should really be part of the definition of the environment of $\mathcal{S}$'' \citep[p. 50]{shaji2005}. This is important because it can be shown \citep[][pp. 13--14]{jordan2004} that the reduced dynamics of a system $\mathcal{S}$ are describable by a completely positive map only if $\mathcal{S}$ is initially not entangled with its environment.\footnote{\citeauthor{jordan2004}'s result generalizes an earlier result for two-dimensional systems proved by \citet[]{pechukas1994}.}

What does it mean, from a physical point of view, for a dynamical map on $\mathcal{S}$'s state space to be not completely positive? It means that a trivial extension of the map to a map on the state space of $\mathcal{S}+\mathcal{W}_n$ is such as to map some states in that larger state space to unphysical states, i.e., to states which predict negative probabilities for the outcomes of certain measurements on $\mathcal{S}+\mathcal{W}_n$ \citep[see][\S 3]{cuffaroMyrvold2013}. It is important, however, to note \emph{which} states in $\mathcal{S}$'s state space result in unphysical states of $\mathcal{S}+\mathcal{W}_n$ when evolved by a not completely positive map on the state space of $\mathcal{S}$ \citep[see][\S 5]{cuffaroMyrvold2013}, because actually it is only the impossible states of $\mathcal{S}$ in a given setup for which this is true. Recall from Section \ref{s:lindblad-st} that when $\mathcal{S}$ is entangled with another system, there is no way for $\mathcal{S}$ to be in a pure state, or more generally in any state that is not a valid partial trace over the entangled state of the overall system. Such states of $\mathcal{S}$ that are impossible in a given setup are ill-described by a not completely positive map; in general a trivial extension of such a map will evolve an impossible state of $\mathcal{S}+\mathcal{W}_n$ to an unphysical state. As for the actually possible states of $\mathcal{S}$ in a given setup,\footnote{These are called the states in the `compatibility domain' of the map (\citeauthor{jordan2004}, \citeyear{jordan2004}, \S 1; \citeauthor{shaji2005}, \citeyear{shaji2005}, p. 52).} in all such cases a positive but not necessarily completely positive map on $\mathcal{S}$ will evolve the state of $\mathcal{S}+\mathcal{W}_n$ to another physically valid state of $\mathcal{S}+\mathcal{W}_n$.

The witness argument aside, there is a deeper, though related, reason for imposing complete positivity. As \citet{raggio1982} put it:

\begin{quote}
A system-theoretic description of an open system has to be considered as phenomenological; \emph{the requirement that it should be derivable from the fundamental automorphic dynamics of a closed system} implies that the dynamical map of an open system has to be completely positive. (p. 435, our emphasis)
\end{quote}

More concretely, let $\Lambda$ be a completely positive trace-preserving map on the state space of a system $\mathcal{S}$ initially in the state $\rho$. Then according to \emph{Stinespring's dilation theorem} \citep[]{stinespring1955},\footnote{Stinespring's theorem is actually more general in that it applies to completely positive maps whether or not they are trace-preserving. Non-trace-preserving maps are useful for characterizing selective operations, such as assessing the result of a measurement \citep[\S 8.2.3]{nielsenChuang2000}. Maps describing the evolution of density operators, however, must be trace-preserving since density operators are defined to be of unit trace.} corresponding to $\rho$ there is a unique (up to unitary equivalence) pure state $| \Psi \rangle \langle \Psi |$ of a larger system $\mathcal{S+A}$ (where $\mathcal{A}$ is called the `ancilla' subsystem), whose dynamics are unitary, and from which we can derive the in general non-unitary dynamics of $\mathcal{S}$.

Thus it is complete positivity which, through Stinespring's theorem, guarantees that we can always derive the dynamical equation for an open system in the way that we did in Section \ref{s:st}, i.e., in terms of a contraction (via the partial trace) of the dynamics of a larger closed system $\mathcal{S+E}$. As \citet[\S 4]{sudarshan2003} point out, a not completely positive map on the state space of $\mathcal{S}$ may also be described in terms of a contraction, but only if we generalize the way we characterize the state space of $\mathcal{E}$.\footnote{We need, in particular, a Hilbert space with an indefinite metric, and we are required to take the evolution of $\mathcal{S}+\mathcal{E}$ to be pseudo-unitary rather than unitary in general. For more on the relation between these concepts, see \citet[]{mostafazadeh2004}. See also \citet[p. 242]{ascoli1958} who show ``that any probabilistically interpretable quantum theory using a Hilbert space of indefinite metric is equivalent to a quantum theory using a Hilbert space of positive definite metric.'' For our purposes the upshot of this is that not every Hilbert space with an indefinite metric is probabilistically interpretable. If the state space of a system is not probabilistically interpretable then it is not physical in the sense in which we have been using that term in this paper.} Unlike the case of a contraction to a completely positive map, however, such contractions (to not completely positive maps) ``have to be viewed as contractions of the evolution of unphysical systems'' \citep[p. 5080]{sudarshan2003}.

This is problematic on the closed systems view, for which the evolution of an open system $\mathcal{S}$ is always described in terms of the evolution of a larger closed system $\mathcal{S+E}$. On the closed systems view, a not completely positive map on $\mathcal{S}$'s state space makes no physical sense, though one might perhaps permit its use as long as one understands that it is merely a mathematical tool \citep[cf.][]{cuffaroMyrvold2013} that does not describe the dynamics of an open system in a fundamental sense. On the open systems view, by contrast, it is not really surprising, and in any case not at all problematic, that the larger closed system from which we may (if we find it convenient to do so) derive a not completely positive map on the state space of $\mathcal{S}$ via a contraction is unphysical. It is neither surprising nor problematic because the methodology of the open systems view does not require that we model an open system as a subsystem of a closed system. On the open systems view there is no need to conceive of the dynamics of an open system as a contraction of anything.\footnote{For more on the properties of not completely positive maps, see \citet[]{dominyEtAl2016, rohiraEtAlConstructionCPMaps}.}

Let us again take stock. We began Section \ref{s:gt} by summarizing the conceptual core of \textbf{GT}: that the evolving state of a physical system, $\mathcal{S}$, is described in that framework in terms of an in general non-unitarily evolving density operator. In Section \ref{s:lindblad-gt} we then showed how to derive, in \textbf{GT}, the same form of the Lindblad equation that one derives in \textbf{ST} for a specific quantum system (see Section \ref{s:st}). In this, final, subsection of Section \ref{s:views-frameworks-qt}, we have seen that there is more to \textbf{GT} than the Lindblad equation. We illustrated this by considering, more carefully than we did in Section \ref{s:lindblad-gt}, one of the important principles assumed in its derivation, namely, the principle of complete positivity. We saw that while relaxing complete positivity makes no physical sense on the closed systems view associated with \textbf{ST}, it is not similarly problematic on the open systems view associated with $\textbf{GT}$.\footnote{One of the reasons for our focusing on complete positivity is that it nicely illustrates an important way in which the closed and the open systems views differ. We note, however, that \citet[]{bassi2013} have shown that complete positivity does not need to be separately assumed in the derivation of the Lindblad equation if the dynamics of a system are assumed to be Markovian. We remark that, as with fundamental not completely positive dynamics, fundamental non-Markovian dynamics are not conceptually problematic on the open systems view either, simply because these are among the possible dynamical evolutions of an open system (in particular one that is strongly coupled to its environment).}

We will revisit these issues in Section \ref{s:ontic-fund}. In the meantime, this completes the expository part of this paper. In the next section we move on to the philosophical issue of interpretation, with a particular focus on the question of how the ontologies of \textbf{ST} and \textbf{GT} actually differ. We will see that despite being formulated in accordance with the closed and open systems views, respectively, the differences between them are not so stark.


\section{Interpreting \textbf{ST} and \textbf{GT}}
\label{s:interpreting-st-and-gt}

The presumed ontology of \textbf{GT} is (comparatively) clear: \textbf{GT} conceives of systems in general as if they were interacting with an external environment; and they are represented by density operators that evolve, in general, non-unitarily over time. But while the question of ontology is, in this sense, comparatively easy to answer in \textbf{GT}, the further question of course arises of what to say about the particular question of the universe as a whole. What does it mean to conceive of the universe as an open system? In fact this is a question of interpretation that, for our purposes at the moment, we do not need to take a position on (we will return to the question, in a sharper form, in Section \ref{s:phil-impl}). We merely remark for now that it is consistent with the methodological presuppositions of \textbf{GT}, as well as with the metaphysical position that motivates them, to conceive of the universe as a closed system. The open systems view does not, \emph{per se}, deny that closed systems exist. It merely denies that a closed system must exist, and represents all systems as, in general, open.

On the closed systems view, the universe as a whole enjoys a privileged ontological status. This is especially true when the parts of the universe are continually interacting with one another, for after all, according to the closed systems view it is in that case the only thing that can truly be described as a system in \textbf{ST}. On the open systems view, by contrast, the universe is just a system like any other, no more ontologically privileged than any of its subsystems, that obeys the same dynamical laws as they do. There is more to say. But for the moment we simply note that already we seem to have said enough to distinguish the ontology of \textbf{GT} from that of \textbf{ST}. As we will see presently, however, this is not the case.

It is instructive to compare the concept of a state in quantum theory with the corresponding concept in classical theory. In classical theory, completely specifying a system's dynamical state yields a simultaneous answer to all of the experimental questions that one can ask about a given observable quantity associated with that system, questions like: \emph{Is the value of the observable quantity $A$ within the range $\Delta$?} In classical theory, once a state assignment is made, the answers to all such questions about the quantity are determined in advance and irrespective of whether we actually ask any questions at all (i.e., irrespective of whether we actually perform an experiment on the system). Further, in classical theory, this is true for \emph{all} observables. That is, once we assign a state to a system, any question concerning \emph{any} of the observable quantities related to it is determined in advance of carrying out whatever process allows one to ascertain the answer.\footnote{\citet[p. 433]{bub-pitowsky2010}, for this reason, refer to classical states as `truthmakers'.}
The same is not true of quantum state descriptions, which suffer from what have been described elsewhere as the `big' and the `small' measurement problems.\footnote{The distinction between a `big' and a `small' measurement problem was first introduced by Itamar \citet[]{pitowsky2006}, and is further developed in \citet[]{bub-pitowsky2010}, \citet[]{bub2016}, and in \citet[]{3m2020}. \citet[]{brukner2017} also distinguishes between a small and a big measurement problem but, unlike these other authors, does not use the terms ironically. Thus Brukner's small problem is these other authors' `big' problem and vice versa. (Otherwise their construals are similar.) Here we follow the exposition given in \citet{3m2020}.}

The big problem (from now on we will omit the inverted commas) is that a given state specification yields, in general, only the probability that the answer to a given experimental question will take on one or another value when asked. This---in itself---is arguably only a relatively minor departure from classical theory, however, since conditional upon the selection of an observable to measure, one can, in quantum theory, describe the observed probabilities (over the values of that observable) as stemming from an imagined prior classical probability distribution over a corresponding dynamical property of the system that \emph{is} determined in advance by the quantum state (analogously to a statistical-mechanical classical state). This is the flip side of the point, which we made earlier, that there is no way to effectively distinguish, in the context of a local measurement, a properly from an improperly mixed ensemble. That is, conditional upon the selection of a particular observable for measurement, the statistics predicted by the reduced density operator characterizing the state of an improperly mixed ensemble will be effectively indistinguishable from the statistics predicted by the density operator characterizing an ensemble that is a proper mixture of systems in eigenstates of the selected observable. This means that, conditional upon selecting that particular observable for measurement, we can effectively use this properly mixed state to characterize the system from that point forward, despite the fact that the system has actually become entangled with the measuring apparatus as a result of its interaction with it. The conditional probabilities associated with the various values of the selected observable are given by the Born \emph{rule}. Note, however, that the theoretical framework of \textbf{ST} does not include any \emph{law} for the dynamical evolution of a system described by a state vector other than the Schr\"odinger equation (or the Dirac equation in the relativistic case). The dynamics that lead to the appearance of one definite outcome of a measurement, rather than another, is not further described by the theory.\footnote{\label{fn:born-rule}This is a feature of quantum theory that is not interpretation-dependent. As we will discuss in more detail shortly, there are two families of interpretations of \textbf{ST} that take it to be complete: orthodox and Everett interpretations. Neither of them posit anything other than unitary dynamics to characterize the dynamical evolution of a state vector. (In the case of Everettian interpretations this is well-known. For the case of orthodox interpretations, see, e.g., \citet[p. 228]{bub2016}; \citet[secs.\ 3--4]{despagnat2001}; \citet[]{howard2004}). So-called `dynamical collapse' interpretations \citep[]{ghirardi2018} are better thought of as alternative theories than as alternative interpretations. They view quantum theory as incomplete, and supplement it with a dynamical mechanism for the non-unitary evolution, not just of the density operator, but of the state vector as well, that is presented as explaining the appearance of definite outcomes of measurements at macroscopic scales \citep[see also the related `transactional interpretation' of][]{cramer1988, kastner2013, kastner2020}. The view called `wavefunction realism' \citep[]{albert1996, north2013} is arguably \citep[see][]{wallace2020} not an interpretation of \textbf{ST} but of a specific theory (non-relativistic quantum mechanics) that can be formulated in that framework but whose lessons do not generalize \citep[see][]{myrvold2015, wallace2020}. For a recent elaboration and defense of wavefunction realism see \citet[]{ney2021}.}

This brings us to the small measurement problem, viz., that the various probability distributions which can, conditional upon a corresponding physical interaction, be used to effectively characterize the various observable quantities associated with a system, cannot be embedded into a global prior probability distribution over the values of all observables as they can be for a classically describable system. In quantum theory one can only say that \emph{conditional} upon our inquiring about the observable $A$, there will be a particular probability distribution that can be used to effectively characterize the possible answers to that question. Conditional upon the selection of a different observable, we will need to use a different probability distribution (over the values of \emph{that} observable) that is in general incompatible with the first.\footnote{Specifically, they will be incompatible whenever the two observables in question do not commute.} This is usually seen as a problem because \textbf{ST}'s unitary description of a measurement interaction does not, in itself, give us an answer to which of these probability distributions we should prefer.\footnote{What we are here calling the small measurement problem is intimately related to (in the sense that it is precisely what is highlighted by) the thought experiment recently suggested by \citet{frauchigerRenner2018}. For some recent commentaries on this and related thought experiments, see \citet[]{brukner2018, bub2018a, bub2021, dascal2020, deBrotaEtAl2020, felline2020a, healey2018, lazaroviciHubert2019}.}

So far nothing we have asserted should be seen as controversial.\footnote{The statements of the small and big measurement problems are only meant to characterize our observational experience in relation to quantum systems; we have said nothing so far about how to interpret this experience in ontological terms. In some Everettian interpretations, for instance, probabilities are grounded in `branch weights' \citep[sec. 3]{saundersEverettProbability}. Other interpretations will have their own ontological stories to tell about how probabilistic phenomena arise. Nevertheless, each of these ways of interpreting quantum theory needs to be able to either explain, or explain away, the significance of the small and big problems in its own terms.} By contrast, consider the following statements: Because the probability distributions over the values of every classical observable associated with a system are determined by the classical state description, independently of whether a physical interaction through which one can assess those values is actually made, we are invited to think of them as originating in the properties of an underlying physical system that exists in a particular way irrespective of anything external, even though there is nothing in the concept of a value, \emph{per se}, that forces us to think of values as originating in this way. This is not the case in quantum theory, where the more complex structure of observables related by it does not similarly invite the inference from the values of observable quantities to the properties of an underlying system in that sense. What is exhibited by the state vector associated with a quantum-mechanical system is not a collection of observer-independent properties, but the structure of and interdependencies among the (unitarily related) possible ways that one can effectively characterize a system in the context of a physical interaction.\footnote{\label{fn:curiel1}This underlying conception of what a system's state represents is, we think, very similar to what \citet[]{curiel2014} has argued is the general conception of a classical state. We will come back to this in note \ref{fn:curiel2} below.} This, in any case, is according to \emph{orthodox interpretations} of quantum theory, which we will have more to say about presently.

There is an ongoing debate over the issue of realism regarding the state vector, where being a realist means that one takes it to represent the observer-independent state of a physical system. In the framework of \citet[]{harrigan2010}, we call an interpretation of the state vector \emph{$| \psi \rangle$-ontic} if it is taken to represent the real state of a physical system in this sense \citep[cf.][p.\ 244]{primas1990}. More concretely, a $| \psi \rangle$-ontic interpretation takes it that for a given real state of the system, $\lambda$, there is a single state vector corresponding to it. If, further, the state vector is taken to \emph{completely} describe the system's real state (i.e., if there is a one-one mapping between $\lambda$ and $| \psi \rangle$), then in addition to being $| \psi \rangle$-ontic such an interpretation is \emph{$| \psi \rangle$-complete}. The various interpretations in the lineage of that of Hugh \citet[]{everett1956}, for instance, are generally regarded as both $| \psi \rangle$-ontic and $| \psi \rangle$-complete, while Bohmian mechanics is a $| \psi \rangle$-incomplete interpretation despite being $| \psi \rangle$-ontic.

In the framework of \citet[]{harrigan2010} one can also describe so-called \emph{$| \psi \rangle$-epistemic} interpretations of the state vector. Such interpretations take the same real state of a system, $\lambda$, to in general be compatible with more than one quantum state. Einstein's interpretation \citep[]{howard1985} of quantum theory is a $| \psi \rangle$-epistemic interpretation in this sense, and so are its modern successors \citep[e.g.,][]{spekkens2007}. Such interpretations are also, clearly, $| \psi \rangle$-incomplete insofar as they take there to be a real state, $\lambda$, that is incompletely described by $| \psi \rangle$.

Orthodox interpretations of the state vector are neither $| \psi \rangle$-ontic nor $| \psi \rangle$-epistemic interpretations in the senses just specified. Yemima \citet[]{benMenahem2017} has called them `radical epistemic' interpretations. This label, though from a certain point of view not inappropriate, is misleading for reasons which should become clearer as we progress.\footnote{Although we think the label `radical epistemic' is misleading, we otherwise do not take issue with Ben-Menahem's very illuminating characterization of orthodox interpretations and explanation of why they fall outside of the scope of the Pusey-Barrett-Rudolph theorem.} For now the important point to note is that, on orthodox interpretations of quantum theory, we are not to think of a system as having a real---in the sense of an observer-independent---state, $\lambda$, at all, for the reasons suggested by our discussion of the big and especially the small measurement problems above.

\begin{figure}
  \begin{center}
    \includegraphics[scale=0.30]{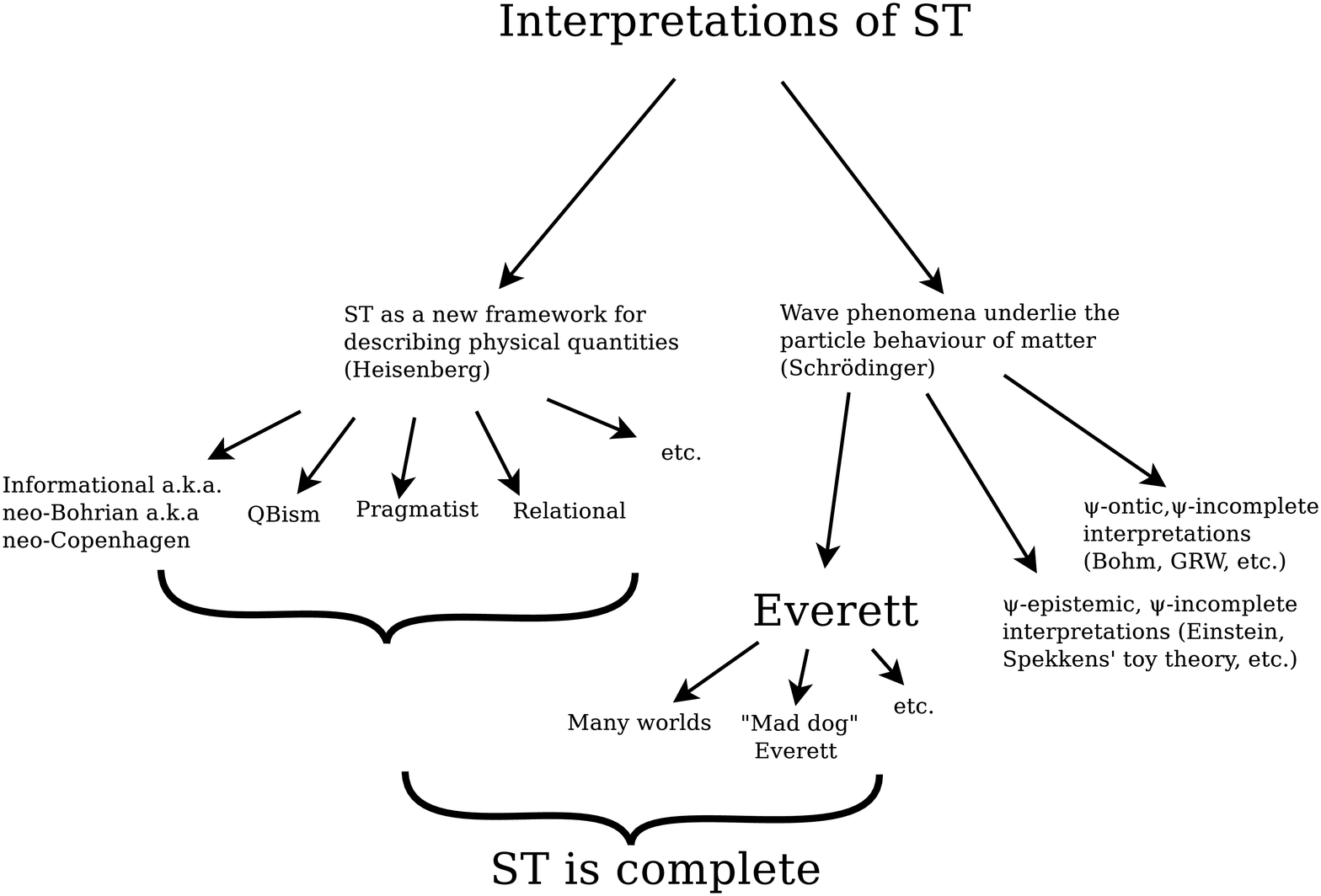}
  \end{center}
  \caption{A genealogy of interpretations of \textbf{ST} \citep*[ch. 1]{3m2020}.}
  \label{fig:genealogy}
\end{figure}

We mentioned the ontological models framework of \citet[]{harrigan2010} because it will be familiar to readers familiar with the Pusey-Barrett-Rudolph theorem \citep[]{pbr2012}, which proves that what one might call `non-radical' epistemic interpretations of a particular sort are impossible.\footnote{For an illuminating discussion of the theorem's assumptions, see \citet{leifer2014}. See also \citet[]{barrettEtAl2013} and \citet{leiferMaroney2013} for related no-go theorems.} As orthodox interpretations of quantum theory cannot be formulated in this framework, however, the ontological models framework turns out to be of rather limited use. A more helpful (though informal) way of relating the various interpretations of quantum theory, given in \citet[ch.\ 1]{3m2020}, is as a genealogy, pictured as a phylogenetic tree with two main branches (see Figure \ref{fig:genealogy}). One of these branches follows, roughly, the lead of Erwin Schr\"odinger, for whom the big discovery of quantum mechanics was that a wave phenomenon underlies the particle behavior of matter. The other, historically older, `orthodox' branch, roughly follows the lead of Werner Heisenberg, for whom the big discovery of quantum mechanics was that solving the problems atomic physics was faced with, in the crisis years of the 1920s, required a new framework for representing physical quantities.\footnote{For more on the genesis of quantum mechanics, see \citet{duncanJanssen2019}.}

Both of these main branches contain further subdivisions. The former branch (the branch on the right in Figure \ref{fig:genealogy}) divides into those interpretations that take quantum theory to be complete, i.e., the various Everettian interpretations which we will have more to say about later, and those that do not, such as Einstein's view as well as the various versions of Bohmian mechanics and dynamical collapse theories. On the other branch we find the various versions of the so-called `Copenhagen interpretation' and their descendants. These also take quantum theory to be complete (though in a different sense). They include, but are not limited to, the connected group of ideas of Niels \citet[]{bohr1928, bohr1935, bohr1949, bohr1958}, Werner \citet[]{heisenberg1959, heisenberg1971}, Grete \citet[]{hermann1935}, Wolfgang \citet[]{pauli1994}, John \citet[]{vonNeumann1932}, Carl Friedrich \citet[]{weizsacker1985}, and others, though they do not constitute a single unified view \citep[see][]{howard2004}. More recent offshoots of this `orthodox' branch include the objectivist neo-Bohrian and neo-Copenhagen interpretations of, for instance, \citet[]{brukner2017, bub2016, bub2017, bub-pitowsky2010, demopoulos2012, demopoulosOnTheories, 3m2020, landsman2006, landsman2017, pitowsky1989, pitowsky2006}; the subjectivist interpretation called QBism \citep[]{fuchsEtAl2014, fuchs2017}; Richard Healey's \citeyearpar[]{healey2012, healey2017} pragmatist interpretation; Carlo Rovelli's \citeyearpar[]{rovelli1996, rovelli2021} relational interpretation, and others.\footnote{Given its radical subjectivity, it may seem strange to call QBism an \emph{orthodox} interpretation, but this is entirely appropriate given our visualization of interpretations of quantum mechanics as a phylogenetic tree. QBism begins with the (objectivist) ideas of Bohr, Pauli, John Archibald \citet[]{wheeler1983}, and others, and then diverges from them \citep[]{fuchs2017}, especially as it regards the interpretation of probability (which for QBists is subjective). As far as we can tell, however, QBism and all of the other views we call orthodox share the very general features that we lay out in this section.}

Orthodox interpretations, which deny that we should think of a system as having a state independently of a given observational context, will truly seem radical for someone who finds it hard to imagine describing reality in an `observer-dependent' way.\footnote{Our use of the term `observer' here should not necessarily be understood in the thick metaphysical sense; i.e., as some agent (embodied or otherwise). Exactly how one interprets the concept of an observer depends on the particular version of the orthodox interpretation being defended. For instance, in \citet[]{3m2020} the concept of an observer is understood schematically \citep[cf.][]{curiel2020}, associated with a particular choice of `Boolean frame'. QBism takes the other extreme, and incorporates a metaphysically thick conception of an agent into their worldview \citep[sec.\ 2]{fuchs2017}. On the relational view an observer is any physical system in interaction with the system of interest \citep[p.\ 1641]{rovelli1996}. On the pragmatist interpretation, an observer is associated with a context (alternately: an `environment') within which the truth value of a given statement can be assessed \citep[sec.\ 3]{healey2021}.} Be that as it may, the so-called ideal of the `detached observer', according to orthodox interpreters, is one that we have found has a limit to its usefulness \citep[cf.][]{nagel1989}. Quantum theory, that is, is construed as placing restrictions, not so much on what we can know (although it does that too) but on what we can conceive of knowing (\citeauthor{pitowsky1994}, \citeyear[pp. 113--114]{pitowsky1994}; cf. \citeauthor{vanFraassen2002}, \citeyear{vanFraassen2002}, p. 3). These restrictions are not absolute (\citealt[]{hermann1933}; \citealt[sec.\ 7]{hermann1935}; \citealt[ch.\ 7]{3m2020}), but they are quantifiable \citep[]{bub2010a, bub2012, cuffaro2018b, cuffaroInMiranda}.
One may choose to call such interpretations epistemic if one wishes, but it is clear that this is a very different sense of that word than the one explicated in the ontological models framework of \citeauthor[]{harrigan2010}.

Despite being `radically epistemic' (in the above sense), defenders of orthodox interpretations claim that they are realist interpretations. However it is important to be clear on what they are realist about.\footnote{The various versions of the orthodox view have been called `participatory realist' \citep[]{cabello2016, fuchs2016}. Peter \citet[]{evans2020} has called the kind of objectivity attributed to quantum theory on some versions of the orthodox view `perspectival objectivity'. Even in QBism, although one interprets probabilities in quantum theory (as one does all probabilities) subjectively, the theory is taken to provide us with an objective method for guiding our action \citep[sec. 2.1]{fuchs2017}.} What the state vector compactly encodes, on such an interpretation, is a collection of unitarily-related conditional probability distributions through which one can effectively characterize a given system in the context of a particular physical interaction.\footnote{\label{fn:curiel2}Compare this with Erik Curiel's \citeyearpar[]{curiel2014} construal (esp. sec.\ 3) of the configuration space of an abstract classical system as encoding a description of its kinematically-possible interactions with other abstract classical systems. See also note \ref{fn:curiel1} above.} These ways of effectively characterizing the system are not themselves (though they may appear to be from within their associated physical contexts) literal depictions of an observer-independent reality, but they are really constrained, and they are constrained in a particular way that is determined by the quantum state vector independently of any particular observer (\citealt[sec.\ 4]{despagnat2001}, \citealt{healey2016}).

As for the system whose effective characterizations, conditional upon a corresponding physical interaction, are really constrained by the state vector, it is in fact conceived of not as a closed but as an open system on an orthodox interpretation. When a system $\mathcal{S}$ is in a state like
\begin{align}
  \label{eqn:superpos-z}
  | \psi_1 \rangle = \alpha \, | b_1^+ \rangle + \beta \, | b_1^- \rangle,
\end{align}
which is unitarily related to the state
\begin{align}
  \label{eqn:superpos-x}
  | \psi_2 \rangle = \alpha' \, | b_2^+ \rangle + \beta' \,| b_2^- \rangle,
\end{align}
and so on, what this signifies is that coupling the degrees of freedom of $\mathcal{S}$ to those of a further system $\mathcal{M}$ will allow one to describe $\mathcal{S}$ in terms of a collection of unitarily-related conditional probability distributions over the possible outcomes of an assessment of $\mathcal{M}$ as described with respect to a particular basis $b_m$. Consider Bohr:

\begin{quote}
In the treatment of atomic problems, actual calculations are most conveniently carried out with the help of a Schr\"odinger state function, from which the statistical laws governing observations obtainable under specified conditions can be deduced by definite mathematical operations. It must be recognized, however, that we are here dealing with a purely symbolic procedure, \emph{the unambiguous physical interpretation of which in the last resort requires a reference to a complete experimental arrangement.} Disregard of this point has sometimes led to confusion, and in particular the use of phrases like `disturbance of phenomena by observation' or `creation of physical attributes of objects by measurements' is hardly compatible with common language and practical definition. \citep[][pp.\ 392--393, our emphasis]{bohr1958}  
\end{quote}

Consider, also, von Neumann:

\begin{quote}
[O]ur knowledge of a system $\mathcal{S}'$, i.e., of the structure of a statistical ensemble $\{ \mathcal{S}'_1, \mathcal{S}'_2,$ $\ldots \}$, is never described by the specification of a state---or even by the corresponding [vector $| \phi \rangle$]; but usually by the result of measurements performed on the system. \citep[p.\ 260]{vonNeumann1927}\footnote{We have slightly changed the mathematical notation to be consistent with our own. For further discussion of this passage see \citet[pp.\ 251--252]{duncanJanssen2013}.}
\end{quote}

Consider, finally, an ensemble of identically prepared entangled pairs; i.e., a number of systems, each of which is composed of two interacting subsystems, $\mathcal{S}$ and $\mathcal{M}$, that are entangled together in the state $| \Psi \rangle_{\mathcal{S}+\mathcal{M}}$. If, as we saw earlier, we now consider the sub-ensemble of this ensemble that is formed by choosing $\mathcal{M}$ from each pair, and inquire of the systems in that sub-ensemble concerning the observable $O$ (i.e., if we make a reading of the measurement instrument), we will find that the resulting statistical distribution of answers will be effectively characterisable as having arisen from experiments on a properly mixed ensemble of systems, each of which is in an eigenstate of the selected observable. Strictly speaking, however, in the context of its physical interaction (with $\mathcal{S}_i$) each $\mathcal{M}_i$ \emph{cannot possibly be} in some state $| \phi_j \rangle$ (with probability $p_j$). Strictly speaking, $\mathcal{S}_i$ and $\mathcal{M}_i$ are open systems which we describe using reduced density operators \citep[p.\ viii]{breuer2007}.

$\mathcal{S}$ is here conceived of as, in reality, an open system. But since \textbf{ST} is formulated in accordance with the closed systems view, conceiving of $\mathcal{S}$ as an open system in this framework requires the introduction of a larger Hilbert space that includes the degrees of freedom of a further system $\mathcal{M}$ that one posits to be coupled with $\mathcal{S}$. This is true regardless of scale. Even when $\mathcal{S}$ is the whole universe, what a state vector describing its state signifies is that, in the context of an interaction between $\mathcal{S}$ and the measurement device $\mathcal{M}$ of a hypothetical supreme being (or anything that could interact with the universe as a whole), this being will be able to characterize the outcomes of that interaction in accordance with a set of unitarily-related conditional probability distributions.\footnote{\label{fn:orthodox-universe}Not everyone who defends an orthodox interpretation of \textbf{ST} will accept that it makes sense to imagine either an actual or hypothetical measurement of the universe as a whole. For an example of an orthodox interpretation that does accept this (or at any rate sees nothing conceptually incoherent in it), see \citet[p. 216]{3m2020}. Whatever one thinks about this issue, it makes no difference to the point we want to make here, which is that systems are interpreted as open on orthodox interpretations whenever it makes sense to model a system at all using \textbf{ST}.} There is no question of taking either the state vector or its generalization, the density operator, to truly represent a closed system of observer-independent properties on this understanding of quantum theory. The closed system described by $| \psi \rangle_{\mathcal{S}}$ is merely abstract. It encodes the abstract possibility space of the system in the context of an interaction with an external environment.\footnote{See notes \ref{fn:curiel1} and \ref{fn:curiel2} above.} The real system represented is an open system, most generally described by a density operator. Open systems, most generally described by density operators, are the stuff of the world on orthodox interpretations of \textbf{ST} and there is no place in \textbf{ST}'s ontology, under this interpretation, for closed systems.\footnote{We note that the statement that open systems are real in \textbf{ST} is compatible with the methodological presuppositions of the closed systems view because, conditional upon a given interaction between $\mathcal{S}$ and $\mathcal{M}$, it is always possible to model the dynamics of $\mathcal{S}$ in terms of a probability distribution over the evolutions, each in accordance with a particular Hamiltonian, of a given set of state vectors.}

Open systems (in addition to closed systems) are also part of \textbf{ST}'s ontology for Everettian interpretations, which comprise the other main family of interpretations of \textbf{ST} whose defenders, like the defenders of orthodox interpretations, take it to be complete. Earlier we discussed the big measurement problem; i.e., the problem raised by the fact that a given state specification yields, in general, only the probability that the answer to a given experimental question will take on one or another value when asked. An advocate of an Everettian interpretation solves this problem by asserting that all of the possible outcomes actually obtain in some sense. This does not solve the small measurement problem, however, which is known to Everettians as the preferred basis problem. The problem, from the Everettian point of view, is that a given state vector can be decomposed in more than one eigenbasis corresponding to the different observables associated with a system. But depending on the eigenbasis chosen we will have a different story to tell about the `worlds' (corresponding to possible outcomes) that are described as existing by the state vector.

For instance a spin-$\frac{1}{2}$ system whose state is expressible as $(| z^+ \rangle ~+~ | z^- \rangle)/\sqrt{2}$ in the eigenbasis of the Pauli $Z$ operator, is expressible in the eigenbasis of the Pauli $X$ operator simply as $| x^+ \rangle.$ Should we take there to be two worlds in this physical situation, or only one? The problem is solved, at least in part, by appealing to the dynamical process of decoherence. When a system, $\mathcal{S}$, interacts with an external environment---for instance some measurement apparatus, $\mathcal{M}$, which is itself coupled to the rest of the universe outside of the lab---then the terms in the superposition describing $\mathcal{S}$'s state decohere \citep[]{zurek2003} and come to achieve a kind of independence from one another when decomposed into a given eigenbasis of the density operator characterizing $\mathcal{M}$. We have seen the effects of this dynamical process already. This is the reason why it is possible to, as we mentioned above, effectively describe a given system $\mathcal{S}_i$, from an improperly mixed ensemble, as if it is in some pure state $| \psi_j \rangle$ with probability $p_j$ even though in reality it cannot be in a pure state since it is entangled with $\mathcal{M}_i$.

How much remains to be solved after taking decoherence into account will depend on the particular Everettian interpretation being defended (a point we will return to shortly). But all Everettians should agree that the point of view, on one part of the universe, $\mathcal{B}$, of an agent who is herself just one part of the rest of the universe, $\mathcal{A}$, which just happens to be so constituted that it functions as a `measuring apparatus' (i.e., it can reliably be correlated with different states of $\mathcal{B}$ either by or for the agent), is itself a matter of fact that is completely described by the state vector for the combined interacting system $\mathcal{A}+\mathcal{B}$. In particular, with respect to a given eigenbasis of the density operator characterizing $\mathcal{A}$, the state vector for $\mathcal{A}+\mathcal{B}$ predicts statistics for measurements on $\mathcal{B}$ that can, effectively, be described as stemming from experiments on members of a hypothetical proper mixture of various proportions, $p_j$, of systems in pure states $| \psi_j \rangle_{\mathcal{B}}$. On the Everettian view, each $| \psi_j \rangle_{\mathcal{B}}$ is then taken to represent the world to anyone who is able to conceive of such a notion and who happens to be a part of what is described by the corresponding state, $| \phi_j \rangle_{\mathcal{A}}$, of $\mathcal{A}$. The world, $\mathcal{B}$, as it is represented by an agent who is a part of $\mathcal{A}$, however, is not the world as described by classical physics. The world for the agent is represented by her as a state vector, a conception of the world that diverges from the classical conception of the world in subtle ways that are only manifested when she considers correlated systems, because correlated systems are in general entangled in quantum theory.

This is true not only of a pair of particles, like the ones considered in the EPR experiment \citep[]{epr1935}, that happen to be at hand to be experimented on by the agent in her world. It is also true of $\mathcal{A}$ and $\mathcal{B}$. To say that $\mathcal{A}$ and $\mathcal{B}$ are entangled is, from one way of looking at the situation, to say that all of the various probability distributions over pure states usable by an agent who is a part of $\mathcal{A}$ to describe the statistics she obtains as she interacts with $\mathcal{B}$ in order to assess the latter's state in the context of various particular measurements, although incompatible with one another if conceived of as all originating in the possessed properties of a single classical object, are nevertheless somehow all consistently encoded in the state vector of the combined system $\mathcal{A}+\mathcal{B}$. This provides some motivation to take them and the worlds (which will, in general, be many) described by each of them to represent something real,\footnote{\label{fn:branch-weights1}Properly speaking what the Everettian takes to be real are the `branch weights', i.e., the squared amplitudes corresponding to the terms in a superposition like the one shown in Eq.\ \eqref{properly-mixed-state}; (objective) probabilities are identified with branch weights \citep[sec. 3]{saundersEverettProbability}.} at least for an agent in $\mathcal{A}$ who reflects on the consequences of her being entangled with $\mathcal{B}$.

As we mentioned earlier, where one goes from here will depend on the particular variant of Everettianism being advocated for. A number of questions will be relevant to this. For instance: Should we take \emph{all} of the worlds so describable to (at least potentially) represent something real \citep[sec.\ 2]{saundersEverettStructure}? Or should we take only those complex enough and stable enough over the course of their lifetimes to be usefully describable, effectively, as a dynamically evolving classical process \citep[]{wallace2003}? Or should something count as a world only if it is approximately describable using the language of classical physics (in the full sense of that term), and can be recognized as such by a dynamical system that happens to be constituted in the right way to play the role of an agent \citep[sec 2.1]{vaidmanSEP}? In fact, the Everett interpretation does not, \emph{per se}, depend essentially on the concept of a world at all \citep[]{carroll2021, carrollSing2018, lehner1997}.\footnote{It has recently been suggested that Jonathan Simon's fragmentalist approach to interpreting the quantum state vector \citep[]{simonFragmenting} is actually ``Everettian Quantum Mechanics in disguise'' or at any rate ``dangerously close'' to it \citep[p.\ 2017]{iaquintoCalosiFragments}. This is incorrect. Simon's fragmentalism is, rather, best construed not as an interpretation but as a meta-interpretation that is intended to be compatible both with unitary-only \textbf{ST} as well as with supplemented versions of \textbf{ST} (i.e., collapse theories) that allow for the possibility of non-unitarily evolving state vectors \citep[sec.\ 3]{simonFragmenting}.}

Regardless of whether one takes all, some, or even none of the classical conditional probability distributions over pure states encoded in the universal state vector to really represent multiverses, there are very good reasons, we think, from the Everettian perspective, to take all such structures to be just as real as anything else described by the universal state vector.\footnote{\label{fn:branch-weights2}As we mentioned above (see note \ref{fn:branch-weights1}), what are real, for Everettians, are not the probabilities \emph{per se}, but the so-called branch weights to which they correspond; i.e., the complex coefficients, or `amplitudes', associated with the terms in a given superposition.} In particular, it is a fact about quantum theory (see Section \ref{s:views-frameworks-qt}) that corresponding to any given probability distribution over pure states of $\mathcal{B}$, one can always find a pure state of some larger system $\mathcal{A}+\mathcal{B}$ from which that probability distribution can be derived.\footnote{For instance, to purify a mixed state, $\rho_{\mathcal{B}} = \sum_ip_i \,| b_i \rangle_{\mathcal{B}}\,_{\mathcal{B}}\langle b_i |,$ where the $| b_i \rangle_{\mathcal{B}}$ are elements of an orthonormal basis for $\mathcal{B}$, we introduce an ancilla system $\mathcal{A}$ which has the same state space as $\mathcal{B}$ and the orthonormal basis $\{\,| a_i \rangle_{\mathcal{A}}\}$. We then define the pure state
  $| \Psi \rangle_{\mathcal{A+B}} = \sum_i\sqrt{p_i} \, | a_i \rangle_{\mathcal{A}} \otimes | b_i \rangle_{\mathcal{B}}.$
  It can be verified that calculating the reduced density operator for $\mathcal{B}$ in relation to this entangled state yields our original density operator $\rho_{\mathcal{B}}$ \citep[][sec.\ 2.5]{nielsenChuang2000}.}

All such `purifications' of the mixed state of $\mathcal{B}$ are, as \citet[][p.\ 171]{darianoEtAl2017} put it, ``essentially the same,'' in the sense that they are all related to one another by a local unitary transformation on the state of the environment; something, in fact, which seems to distinguish quantum theory from other possible physical theories.\footnote{\citet[ch.\ 1]{darianoEtAl2017} list six principles: causality, local discriminability, perfect discriminability, ideal compression, atomicity of composition, and purification, that can be used to characterize theories of (syntactic) information. Classical theory (as applied to information) satisfies the first five. Quantum theory is the only such theory that satisfies all six \citet[pt.\ IV]{darianoEtAl2017}.} Thus the density operator, $\rho_{\mathcal{B}}$, decomposed mathematically into a decoherent mixture of various states corresponding to the elements of an eigenbasis of $\rho_{\mathcal{A}}$, is just as objective a description of everything there is, relative to the degrees of freedom included in our representation of $\mathcal{B}$ and to the given eigenbasis, as the universal state vector $| \Psi \rangle_{\mathcal{A}+\mathcal{B}}$. Further, if we would like to describe the mixed state of $\mathcal{B}$ relative to some other basis, we can (equivalently) either unitarily transform $| \Psi \rangle_{\mathcal{A}+\mathcal{B}}$ and then trace over $\mathcal{A}$, or unitarily transform $\rho_{\mathcal{B}}$ directly by applying the Liouville-von Neumann equation. For the Everettian there can be no reason that one can glean from the framework of \textbf{ST} \emph{per se} to think of an open system, i.e., a system whose degrees of freedom are coupled with those of its environment, whose state is represented by a density operator $\rho_{\mathcal{B}}$ decomposed in a particular basis---regardless of whether one's version of Everettianism licenses taking a particular decomposition of the density operator to represent a multitude of worlds or not---as any less an element of reality than the closed system represented by the universal state vector $| \Psi \rangle_{\mathcal{A}+\mathcal{B}}$ (cf.\ \citealt[secs.\ 8.4, 8.6]{wallace2012}, \citealt[]{wallaceTimpson2010}).

There are no other options for someone who takes \textbf{ST} seriously as a candidate fundamental theoretical framework, i.e., who either interprets it to offer us a complete description of reality (Everettian), or as complete in the sense of providing us with all of the resources we need to describe, to whatever level of detail we like, any given observable physical phenomenon (orthodox). And regardless of whether one favors Everettian or orthodox interpretations, we have seen that there are good reasons to think of open systems, represented by density operators evolving, in general, non-unitarily, as real within \textbf{ST}, just as they are in \textbf{GT}.

Recall from Section \ref{s:intro} that for Schaffer, the lesson of \textbf{ST} is that the cosmos is the one and only fundamental thing. We have seen in this section, however, that despite being formulated in accordance with the closed systems view, the ontology of \textbf{ST} actually includes open systems either exclusively (for orthodox interpreters) or in addition to closed systems (for Everettians). Moreover in the next section we will see that that there is a clear motivation, that arises from within \textbf{ST} and regardless of how one interprets it, to embrace \textbf{GT}---a framework motivated by the metaphysical position that takes open systems to be fundamental---as a more fundamental theoretical framework than it.

\section{The Fundamentality of the Open Systems View}
\label{s:fundamentality}

We saw in Section \ref{s:views-frameworks-qt} that there are two alternative ways to characterize the physics of quantum systems. The first way uses \textbf{ST}, a theoretical framework formulated in accordance with the closed systems view, and represents closed systems using state vectors that evolve unitarily. The second way uses \textbf{GT}, a theoretical framework formulated in accordance with the open systems view, which represents all systems using density operators that evolve, in general, non-unitarily. The question that we will be addressing in this section concerns which of these characterizations is the more fundamental one. We will consider three alternative notions of fundamentality as it relates to a comparison between theoretical frameworks in the course of our discussion. One of these is \emph{ontic fundamentality}, which relates to the objects described by a theoretical framework.\footnote{We will explain what we mean by an object of a theoretical framework in the next subsection.} The second is \emph{epistemic fundamentality}, which relates to our knowledge of those objects. The third is \emph{explanatory fundamentality}. In the remainder of this section we will argue that \textbf{GT} is a more fundamental theoretical framework than \textbf{ST} regardless of which of these ways of cashing out fundamentality one adopts.


\subsection{Ontic Fundamentality}
\label{s:ontic-fund}

In Section \ref{s:views-frameworks-qt} we distinguished between models, theories, and theoretical frameworks. We begin this section by noting that the question of ontic fundamentality is relevant to all three of these conceptual levels. In the context of particle physics one might say, for instance, that a model of a phenomenon given in terms of quarks is more fundamental than one given in terms of protons, appealing to the fact that protons are `made up' of quarks (and gluons) according to our best theory.\footnote{This is currently referred to as the `standard model,' which is actually not a model in the sense (generally used in the philosophy of science) discussed in Section \ref{s:views-frameworks-qt}, but a theory.} Particle physics is not special in this. In the context of arithmetic, to take a different example, one might claim that integers are more fundamental than rational numbers, given that a fraction is formed by relating two integers and made up of them in that sense \citep[cf.][p.\ II]{frege1980}.

The second example illustrates that the question of relative ontic fundamentality as it relates to two models of a target system in the context of a given theory can be a subtle one.\footnote{In this case we can think of the `target system' as the particular quantity described by the expressions `3', `1/3 + 8/3', `2/3 + 7/3', and so on. What one should make of the ontological status of such target systems is, of course, another matter.} Every integer is the sum of two rationals, after all, thus one could (conceivably) argue on that basis that rationals, not integers, are more fundamental in arithmetic. Note that the above arguments all assume that if $A$ is `made up of' $B$ (whatever that means in a given context\footnote{As Karen \citet[sec.\ 2.1]{bennettbook} would put it, there are multiple `building relations' (which form a unified family) in virtue of which we can take one thing to be more fundamental than some other thing.}), then $B$ is more fundamental than $A$. That is only one way to flesh out ontic fundamentality in the context of a given theory, however. There may be others. But we will set this question aside for now. The important thing to take away from these examples, for our purposes, is that in each case what is being compared are particular models as described within a single theory.

The question of whether one theory or theoretical framework is more fundamental than another is not like the question of whether quarks are more fundamental than protons in particle physics, or whether integers are more fundamental than rational numbers in arithmetic. Answering the question of which of two given theories or theoretical frameworks is more fundamental, where each, within itself, supports its own relations of ontological fundamentality with respect to the objects it describes, can be a subtle and difficult exercise. As Kerry \citet[\S 4]{mckenzie2019} notes, some such questions will be easier to answer than others, at least \emph{prima facie}. A compelling case, for instance, can be made that classical theory is less fundamental than quantum theory because certain aspects of the appearance of classicality can be seen to follow from the unitary dynamics of quantum-mechanical systems via the dynamical process of decoherence. How one interprets this dynamical account of the appearance of classicality is another matter, of course, but putting interpretive questions to one side for the moment, the reason this case is particularly compelling is that (as McKenzie notes) all one needs to do to get the appearance of classicality is to consider the effective dynamics of quantum systems entangled with their external environment.

The above example of the relation between quantum and classical theory seems to suggest that we can call one theoretical framework more fundamental than another one if objects as described by the latter can always be seen to arise as a special case of some aspect of objects as described by the former.\footnote{In the rest of the section we will focus on comparisons between frameworks rather than theories, though we take our discussion to be applicable at the latter level as well.} The basic idea here is to think of fundamentality comparisons between frameworks in terms of a kind of determination (or what Bennett calls a \emph{building}) relation; i.e., it is to think of the more fundamental as somehow being responsible for, i.e., as a \emph{sufficient condition} for bringing about the less fundamental \citep[sec.\ 3]{mckenzie2019}.\footnote{Bennett's particular characterization of building is as a directed relationship between fundamental and non-fundamental such that the former is understood as necessitating and generating the latter \citeyearpar[sec.\ 3.1]{bennettbook}. Both necessitation and generation express sufficient conditions insofar as a particular state of the non-fundamental is taken to be entailed (in one of those senses) by a particular state of the fundamental. Other philosophers who cash out fundamentality in terms of determination include \citet[]{fine2015} and \citet[]{dasgupta2015}.} Motivated by this idea, we now proceed to provisionally explicate the relation of ontic fundamentality between two theoretical frameworks. First, we informally explicate, and then elaborate on, the idea of an \emph{object of a theoretical framework}:

\bq
\textbf{(Object)} Let $\mathcal{F}$ be a theoretical framework. $\mathcal{O}$ is an Object (`capital-O') of $\mathcal{F}$ iff objects (in the `little-o', or `pre-theoretic' sense) are representable in terms of $\mathcal{O}$ in every model of every theory that can be formulated in $\mathcal{F}$.
\eq

Clearly this explication is very informal, but it is enough to convey the idea that the Objects of a theoretical framework are always available irrespective of any particular modeling assumptions one might employ in a given context (other than the minimal assumptions required by the framework). In other words, the Objects of a theoretical framework codify its methodological presuppositions; i.e., the ways in which (little-o) objects may be modeled in the framework. Let us illustrate this in more concrete terms using the examples of \textbf{ST} and \textbf{GT}. Consider: (i)~In every model of every theory that can be formulated in \textbf{ST}, the state of a closed system, $\mathcal{S}$, is given by a unitarily evolving normalized state vector, $| \psi \rangle$; (ii)~in every model of every theory that can be formulated in \textbf{GT}, the state of a system, $\mathcal{S}$, is given by an in general non-unitarily evolving density operator, $\rho$. Our little-o object is $\mathcal{S}$ in each case.\footnote{Our understanding of a little-o object is similar to the way that \citet[]{primas1990} characterizes a system: ``By a \emph{system} we just mean the referent of a theoretical discussion'' (p.\ 244). Note, however, that our understanding of an Object (capital-O) does not correspond to what Primas calls an object: ``An object is defined to be an open quantum system, interacting with its environment, but which is not Einstein-Podolsky-Rosen-correlated with the environment'' (ibid.).} Our Objects are $| \psi \rangle$ and $\rho$, by which we do not mean any particular instantiation of $| \psi \rangle$ or any particular instantiation of $\rho$ but the \emph{abstract concepts} of a state vector and of a density operator in general, instances of which we draw upon to represent a given system in a given model. From here on, whenever we discuss the `objects of a theoretical framework' (or use a similar expression), what we mean are objects in the sense of Object, but we will generally not use (except in cases where it makes sense to do this for emphasis) a capital-O.\footnote{Note that our distinction between little-o and capital-O objects is similar to \citeauthor[]{ladymanRoss2007}'s \citeyearpar[sec. 2.3.4]{ladymanRoss2007} distinction between what they call (following Carnap) the formal and material modes of speech.}

Next, consider the following explication of ontic fundamentality as it relates to two objects of a given framework:
  
\bq
{\bf (OntFund-O)} Let $\mathcal{O}_F$ and $\mathcal{O}_P$ be any two objects of a given theoretical framework, $\mathcal{F}$. $\mathcal{O}_F$ is \emph{ontologically more fundamental} than $\mathcal{O}_P$ with respect to $\mathcal{F}$ iff whenever an instance of $\mathcal{O}_P$ appears in any model of any theory that can be formulated in $\mathcal{F}$, some instance of $\mathcal{O}_F$ can be understood to determine $\mathcal{O}_P$ in that model. Furthermore, $\mathcal{O}_F$ is \emph{ontologically fundamental} in $\mathcal{F}$ iff nothing is more fundamental than $\mathcal{O}_F$ in $\mathcal{F}$.
\eq

We now use \textbf{OntFund-O} to define:

\bq
{\bf (OntFund-1)} Let ${\cal F}_F$ and ${\cal F}_P$ be two theoretical frameworks, and let $\mathcal{O}_P$ be any one of the fundamental objects (in the sense of \textbf{OntFund-O}) of ${\cal F}_P$. ${\cal F}_F$ is \emph{ontologically more fundamental} than ${\cal F}_P$ iff $\mathcal{F}_F$ always re-describes any instance of $\mathcal{O}_P$ as effectively determined given an instance of some other object, $\mathcal{O}_F$, that is more fundamental in ${\cal F}_F$. Furthermore, $\mathcal{F}_F$ is \emph{ontologically fundamental} iff there is no theoretical framework more fundamental than $\mathcal{F}_F$.
\eq

Note that we have not explicated what it means, in general, to `re-describe' (an instance of) an object of one theoretical framework within some other theoretical framework. We do not explicate what this means in general because we do not presuppose that it means the same thing irrespective of the nature of the two frameworks under consideration. We do know, however, what it means in the context of a comparison of \textbf{ST} and \textbf{GT}. As we discussed in Section \ref{s:views-frameworks-qt}, any state vector $| \psi \rangle$ can alternately be described as a density operator by taking the projection onto that state vector: $\rho = | \psi \rangle\langle \psi |$. This density operator describing the `pure' state of a system is equally valid in both \textbf{ST} and \textbf{GT}, while the state vector $| \psi \rangle$ is exclusive to \textbf{ST}. While, as we saw, the way that we characterize the dynamics of systems in \textbf{GT} differs from the way that we characterize them in \textbf{ST}, the density operator functions as a bridge-concept that allows us to more easily compare the two frameworks with one another.

Unfortunately, it seems that \textbf{OntFund-1} cannot help us to determine whether \textbf{ST} or \textbf{GT} is more fundamental. On the one hand, the fundamental (unitary) dynamical evolution of a state vector, $| \psi \rangle$, as described in \textbf{ST} is always re-describable in \textbf{GT} in terms of the fundamental (in general non-unitary) dynamical evolution of a density operator, $\rho$. On the other hand, Stinespring's theorem (see Section \ref{s:wider-framework}) shows us how to derive the non-unitary evolution of a given density operator, $\rho_{\mathcal{S}}$, from the fundamental unitary evolution of a state vector $| \Psi \rangle_{\mathcal{S+E}}$. Note that the failure of \textbf{OntFund-1} has nothing to do with the fact that it is motivated by the relation of determination rather than, say, dependence (according to which the more fundamental is thought of as a necessary rather than sufficient condition for the less fundamental \citep[p.\ 57--58]{mckenzie2019}). Since we can always translate \textbf{ST}'s picture of an evolving system into the language of \textbf{GT} and vice versa without any loss of information, it should be clear that neither dependence nor determination can help.

It might be thought that we can break the stalemate by recalling (see Section \ref{s:wider-framework}) that \textbf{GT} has greater expressive power than \textbf{ST} in the sense that dynamical maps on density operators that do not make physical sense in the latter framework have a natural interpretation in the former. But unless we independently motivate these dynamical possibilities, appealing to the greater expressive power of \textbf{GT} to argue for its relative fundamentality is circular insofar as it is essentially to appeal to the very dynamical possibilities that the advocate of the closed systems view (from which \textbf{ST} is formulated) would seem to find objectionable. We will come back to this later. In the meantime we will restrict our attention to the dynamical descriptions that can be made sense of in both frameworks.

The trouble with \textbf{OntFund-1} is that it is too abstract. In particular we have not considered the little-o objects, i.e., the ontologies, that the capital-O objects of \textbf{ST} and \textbf{GT} actually represent. Recall from Section \ref{s:interpreting-st-and-gt} that in regard to this question, \textbf{GT} was found to be clear in at least this respect: Its fundamental objects represent systems as open in general. As for \textbf{ST}, we saw that characterizing its ontology is somewhat more complicated, but that if one takes \textbf{ST} to be a candidate fundamental theoretical framework, then there are only two families of interpretational options available: Everettian and orthodox; and on either option one should conclude that \textbf{ST} is about open systems (at least in part) despite being formulated in accordance with the closed systems view.

Recall from Section \ref{s:views-frameworks-qt} that what we mean when we say that \textbf{ST} is formulated in accordance with the closed systems view is that, within \textbf{ST}, because the Hamiltonian associated with a system does not include terms to represent its interaction with its environment, modeling the dynamics of an open system, $\mathcal{S}$, involves showing how they arise as a result of the coupling between $\mathcal{S}$ and a further system $\mathcal{E}$ that together form one closed system. And what we mean when we say that \textbf{ST} is about open systems despite this is that, as we saw in Section \ref{s:interpreting-st-and-gt}, its little-o objects are open systems (exclusively) for the orthodox interpreter, and include both open and closed systems for the Everettian. \textbf{GT}, by contrast, is formulated in accordance with the open systems view. On this view we do not model the influence of the external environment on $\mathcal{S}$ in terms of an interaction between two systems. We rather represent that influence in our formal description of the dynamics of $\mathcal{S}$. In the remainder of this subsection, we will argue that whether one favors an Everettian or an orthodox interpretation---or neither---there are good reasons to take \textbf{GT} to be ontologically more fundamental than \textbf{ST}.

On an orthodox interpretation of \textbf{ST}, the closed system represented by a state vector is merely abstract; it codifies a collection of unitarily-related conditional probability distributions through which one can effectively characterize a given open system in the context of a physical interaction.\footnote{See notes \ref{fn:curiel1} and \ref{fn:curiel2}.} Open systems, which are most generally represented by density operators, are the stuff of the world on orthodox interpretations of \textbf{ST}. There is no room for closed systems, in an ontological sense, at all; not even for the closed system of interrelated observational possibilities represented by the universal state vector. But since \textbf{ST} is formulated in accordance with the closed systems view, expressing that $\mathcal{S}$ is an open system, regardless of whether $\mathcal{S}$'s state is given in terms of a density operator or a state vector, even in the case where what is being described is the entire physical universe,\footnote{See note \ref{fn:orthodox-universe}.} \emph{always} involves introducing a hypothetical second system in interaction with the first, both of which are conceived of as open systems.

Thus the picture of a physical system that is presented to us by \textbf{GT}, which describes it as an in principle non-unitarily evolving open system, is a picture of a physical system that an orthodox interpreter is already ontologically committed to on the basis of her understanding of \textbf{ST}. \textbf{GT} differs from \textbf{ST} insofar as it strips away the ``ultimate observer'' from our description of a given quantum system \citep[cf.][sec.\ 4]{bub2018a}; i.e., the second system that we introduce, in any theoretical framework formulated in accordance with the closed systems view, to account for the dynamics of a given open system. But despite this an orthodox interpreter's underlying conception of a system as open need not change in the move from \textbf{ST} to \textbf{GT}. That is, although the degrees of freedom associated with the observer are stripped away from our description of a system in \textbf{GT}, the density operator characterizing the system may straightforwardly be interpreted in the same way as before; i.e., as encoding a probability distribution over the ways that the system will manifest itself to a user of the theory that is interacting with it. Finally, \textbf{GT} makes fully explicit what the conditions are that are required in general to recover the dynamical evolutions that are possible in \textbf{ST}. The orthodox interpreter thus has, we conclude, very good reasons to embrace \textbf{GT} as the more fundamental theoretical framework.

Let us now turn to Everettian interpretations. We begin by considering more carefully the ways in which \textbf{GT} and \textbf{ST} actually differ. For the time being we leave aside the difference in expressive power between these two frameworks that relates to the issue of the principle of complete positivity, and continue to restrict our attention to the dynamical descriptions that make sense in both frameworks. With respect to this overlap area, \textbf{GT} and \textbf{ST} are, evidently, predictively equivalent. Predictive equivalence on a given domain does not imply empirical equivalence (let alone theoretical equivalence) on that domain, however. In particular, \citet[]{curiel2014} has shown how two predictively equivalent theoretical frameworks can evince empirically significant differences when we examine the global features of each.\footnote{\label{fn:weatherall}For a discussion of some different senses of empirical inequivalence (including those that appeal to inequivalence in expressive power) and of theoretical equivalence more generally, see \citet[]{weatherall2019a, weatherall2019b}.} One of the global features of a theoretical framework consists in the way that it allows one to describe everything in its domain, which in this case (since these are frameworks for physical theories) comprises everything that physically exists. In the case of \textbf{GT}, the picture one draws of the physical universe's evolving state describes it as an in principle non-unitarily evolving density operator, while in the case of \textbf{ST} one depicts it in terms of a unitarily evolving state vector.

Let us then ask ourselves the question: \emph{Must} the universe as a whole be described in terms of a unitarily evolving state vector for an Everettian? The answer is no, at least according to some Everettians.\footnote{For an example of an Everettian who does require this, see \citet[]{deutsch1991}.} David \citet[sec.\ 10.5]{wallace2012}, for instance, argues quite explicitly that there are no philosophical reasons (for an Everettian or otherwise) not to take seriously the possibility that the universe's dynamics are those of a non-unitarily evolving density operator (pp.\ 397--400), and further, that there is positive evidence for this possibility that comes from considering the physics of black holes (pp.\ 400--401). Wallace \citep[see][]{wallace2020a} no longer finds that particular evidence convincing. But regardless of what one makes of this and other evidence, the point we want to make here is that there is no a priori reason to think that describing the universe in terms of a non-unitarily evolving density operator is somehow inconsistent with the Everett interpretation.

As an illustration, consider the simple density operator below, expressed as a mixture of two pure states:
\begin{align}
  \label{eqn:mixed}
  \rho = p \, | \psi_1 \rangle \langle \psi_1 | \;+\; (1-p) \, | \psi_2 \rangle \langle \psi_2 |.
\end{align}
The fact that the different terms, $| \psi_1 \rangle \langle \psi_1 |$ and $| \psi_2 \rangle \langle \psi_2 |$ are by definition \emph{decoherent} makes it unproblematic, irrespective of whether $\rho$ evolves unitarily, to identify them with independently evolving worlds. In other words, unlike the case of a unitarily evolving state vector, where we must appeal to environmental decoherence as well as to pragmatic considerations in order to argue that, \emph{for all practical purposes}, the worlds that one identifies in the associated superposition are effectively independent of one another even though, strictly speaking, they are not \citep[]{wallace2003}; in the case of a literally decoherent mixture like the one described in Eq. \eqref{eqn:mixed}, the worlds represented by $| \psi_1 \rangle \langle \psi_1 |$ and $| \psi_2 \rangle \langle \psi_2 |$ are independent in a strict sense and not just independent for all practical purposes. Thus the motivation to think of the quantum description of the universe as consisting in a description of multiple independently evolving worlds is even more clear on \textbf{GT}, where a fundamental description of the universe may take the form of \eqref{eqn:mixed}, than it is on \textbf{ST}.

Sean Carroll, in a paper entitled ``Reality as a Vector in Hilbert Space'' concedes that: ``Technically [the state of the universe] is more likely to be a mixed state described by a density operator, but that can always be purified by adding a finite-dimensional auxiliary factor to Hilbert space, so we won't worry about such details'' \citep[p. 5]{carroll2021}. Carroll, unlike Wallace, does not comment on whether the mixed state of the universe should be understood as evolving non-unitarily or unitarily. For our part we see no reason to rule out the possibility that it could be evolving non-unitarily, for it is simply the case that density operators evolve, in general, non-unitarily in \textbf{ST}. If we are to take seriously the idea that the universal state is represented by a density operator (as Carroll plainly does), then we should take the possibility that it is represented by a non-unitarily evolving density operator just as seriously. And if we do so, then we should view \textbf{GT} as more fundamental than \textbf{ST}, since the former is the only framework that actually permits us to model the state of the universe fundamentally in terms of a non-unitarily evolving density operator.

Of course, nothing we have just said constitutes a reason for thinking that a description of the universe given in terms of a non-unitarily evolving density operator is any more likely to be true than a description in terms of a unitarily evolving state vector. But it should be clear, we think, that both descriptions are perfectly consistent with Everettianism. Ultimately, for an Everettian, which of \textbf{GT} or \textbf{ST} one adopts should not be decided upon dogmatically. The answer should rather depend on the physical evidence. Given the \emph{prima facie} evidence and conceptual reasons we discussed in Section \ref{s:intro}, we think there is every reason to at least take the former possibility---the idea that the universe's evolution takes the form of the evolution of an open system---completely seriously in fundamental physics; i.e., as a live option that any theoretical framework should allow us to make sense of. And if we do this then there is every reason to embrace \textbf{GT}, rather than \textbf{ST}, as our preferred theoretical framework for quantum theory, given that the former is the only framework that actually permits us to model the dynamics of the universe fundamentally in these terms.

As far as interpretations of \textbf{ST} go, this leaves only hidden-variable theories, which will require only a brief discussion. Hidden-variable theories of \textbf{ST}, given that they are not competing interpretations of \textbf{ST} but different theoretical frameworks entirely, are not really relevant to our comparison between \textbf{GT} and \textbf{ST}. Moreover, since \textbf{GT} shares those features of \textbf{ST} which lead hidden-variable proponents to seek an alternative theoretical framework in the first place---measurement outcomes are still irreducibly probabilistic in \textbf{GT}, and the probability distributions over measurement outcomes for different experiments performable on a system are still in general incompatible\footnote{In Section \ref{s:interpreting-st-and-gt} we called these the big and the small measurement problems respectively.}---those who favor hidden-variable theories are unlikely to embrace \textbf{GT} as a fundamental theoretical framework for physics. There is nevertheless something that advocates of hidden-variable theories may take away from our comparison between \textbf{GT} and \textbf{ST}: that, since there are good reasons to conclude that \textbf{GT} is ontologically more fundamental than \textbf{ST}, regardless of how one approaches the question of interpreting \textbf{ST} under the assumption that it is a candidate fundamental theoretical framework, then the proponent of a hidden-variable interpretation would do better to focus her intellectual energies on \textbf{GT} rather than \textbf{ST} in her efforts to recover whatever she requires of a framework that it be able to do.

For instance, it is possible (though it is by no means required) to formulate a dynamical collapse theory within the framework of \textbf{GT}: Collapse theories \citep[]{ghirardi2018} assume that state vector collapse occurs in position space, and the resulting dynamics can be derived from a suitable Lindblad equation. Such solutions to the measurement problem are of course controversial and they run into problems with a relativistic extension. However an advocate for \textbf{GT} is not obliged to accept collapse theories. It is merely possible to formulate them within this framework. We emphasize, again, that \textbf{GT} is a theoretical framework, not a particular theory. Moreover it is (as we argued in Section \ref{s:wider-framework}), a framework that allows for the modeling of more general dynamical evolutions of a system than \textbf{ST}, and thus provides a richer conceptual landscape with which to explore different responses to the measurement problem and other conceptual puzzles related to quantum theory.

We have now argued: (i)~that open systems are fundamental in \textbf{ST}, regardless of how one interprets \textbf{ST} as a candidate fundamental theoretical framework; (ii)~that, regardless of how one interprets \textbf{ST} \emph{tout court}, one should conclude that \textbf{GT} is more fundamental than \textbf{ST} with respect to the dynamical descriptions that can be made sense of in both. This is enough to conclude that \textbf{GT} is ontologically more fundamental than \textbf{ST}, and thus enough to have fulfilled the goal that we set for ourselves in this subsection. What remains is to provide our proposal, finally, for an explication of ontic fundamentality as it relates to theoretical frameworks:

\bq
{\bf (OntFund-2)}
Let ${\cal F}_F$ and ${\cal F}_P$ be two theoretical frameworks, and let $\{\mathcal{O}_F\}$ and $\{\mathcal{O}_p\}$ be their corresponding fundamental objects (in the sense, say, of \textbf{OntFund-O}).\footnote{We do not think that \textbf{OntFund-2} crucially depends on our cashing out the relation of ontic fundamentality between a framework's objects in terms of determination rather than, say, dependence.} Furthermore let ${\cal F}_F$ and ${\cal F}_P$ be motivated by two distinct metaphysical positions, ${\cal M}_F$ and ${\cal M}_P$, respectively, in regard to their little-o objects. ${\cal F}_F$ is \emph{ontologically more fundamental} than ${\cal F}_P$ iff the way that the $\{\mathcal{O}_p\}$ actually represent their little-o objects corresponds not to ${\cal M}_P$ but to ${\cal M}_F$. Furthermore, $\mathcal{F}_F$ is \emph{ontologically fundamental} iff there is no theoretical framework more fundamental than $\mathcal{F}_F$.
\eq

The fact that the ontological implications of taking a view in a given domain might become decoupled from the metaphysical position that motivates that view in the first place might seem paradoxical, but we can begin to understand how this situation arises in quantum theory when we consider that the closed systems view from which \textbf{ST} is formulated is the view of the world that it inherits from classical (i.e., 19th century) physics. \textbf{ST} generalizes probabilistically (as compared with classical theory) the way in which one represents the state of a closed system while remaining consistent with the methodological presuppositions of the closed systems view. As we have seen, however, the result of this generalization is that what is actually represented by the objects of \textbf{ST} is a domain of little-o objects that includes open systems essentially.

Earlier we noted that \textbf{OntFund-1}, our provisional explication of relative ontic fundamentality between two given frameworks, is unable to answer the question of whether \textbf{ST} is more fundamental than \textbf{GT} with respect to the kinds of dynamical descriptions that can be made sense of in both. At the time we dismissed as circular the attempt to break the deadlock by appealing to the greater expressive power of \textbf{GT}, given that this is essentially to appeal to the very dynamical descriptions (such as those given by a not completely positive map on a system's state space; see Section \ref{s:wider-framework}) that the advocate of the closed systems view (from which \textbf{ST} is formulated) would seem to find objectionable.

The reason why the dynamical evolutions described by, for instance, not completely positive maps on the state space of an open system do not make physical sense within \textbf{ST}, is that \textbf{ST} is formulated in accordance with the closed systems view, which is motivated by the metaphysical position according to which the subject matter of science is isolated systems, and whose methodological presuppositions require that we model the dynamics of an open system in terms of an interaction between two subsystems that together are closed. Our discussion of the various interpretations of \textbf{ST} has shown us, however, that in principle it is possible for the conception of its ontology implied by a theoretical framework to become decoupled from the metaphysical position that motivates it. We have seen in particular that, despite being formulated in accordance with the closed systems view, \textbf{ST} is actually about open systems either exclusively (on an orthodox interpretation) or at least in addition to closed systems (on an Everettian interpretation).

Unless one demands, pre-theoretically, that every open system be understood as a subsystem of a larger closed system---i.e., unless one takes the closed systems view to have, not just methodological force but ontological force as well---then there are good reasons to motivate taking any linear map that evolves one valid density operator to another valid density operator, even a not completely positive map, to have physical significance whether or not that significance is actually expressible, in fundamental terms, in \textbf{ST}. That such maps are not possible to make physical sense of in \textbf{ST} should not, \emph{per se}, be a reason to dismiss them as unphysical \emph{tout court}. It should, rather, be a reason to formulate a more fundamental framework, which might in principle involve a change in corresponding view, from which their physical significance can be understood. In other words there is a clear motivation, that arises from within \textbf{ST} itself, to seek for more general ways to characterize the dynamics of density operators than are possible in that framework. These more general ways of characterizing open systems evolution are what are provided by \textbf{GT}.

\subsection{Alternative Forms of Fundamentality}
\label{s:alternative-funds}

The foregoing arguments are in part premised on our entertaining the possibility that the universe's dynamical evolution might actually take the form of a non-unitarily evolving density operator. But, one might object, why should we believe this? Should we not, rather, take the empirical success of \textbf{ST} to motivate us to believe that the universe is best described in terms of a unitarily evolving state vector? We think not. This objection, on our view, wrongly represents what the basis of the empirical success of \textbf{ST} actually is; for the basis of the empirical success of \textbf{ST} lies in its applications to systems that we actually have empirical access to, i.e., the subsystems of the universe, and although it is often possible to successfully model these as closed, in actual fact they never are (see Section \ref{s:intro}). It is of course true that one can successfully provide what, in the context of \textbf{ST}, counts as an effective description of the dynamics of a given open system, $\mathcal{S}$, by first modelling the dynamics of the closed system $\mathcal{S+E}$, where $\mathcal{E}$ refers to the system's environment, and then abstracting away from (`tracing out') the degrees of freedom of $\mathcal{E}$. But it is typically \emph{not} the dynamics of $\mathcal{S+E}$ which we take ourselves to have successfully described when we do so (this description will typically be very highly idealised); it is, in fact, the dynamics of $\mathcal{S}$.\footnote{Thanks to Wayne Myrvold for suggesting this argument in conversation.} The basis of the empirical success of \textbf{ST}, therefore, lies in the way that it describes the dynamics of open systems. And since the dynamics of an open system, as formally described by \textbf{ST}, is in general non-unitary; there is, we take it, a clear motivation to consider adopting a more general theoretical framework in which it is possible to fundamentally describe the dynamics of systems in that way.

In \textbf{ST}, the unitarily evolving state vector representing the dynamical evolution of the universe as a whole is the single, not directly accessible, exception; it represents the one and only unitarily evolving thing. In contrast, in \textbf{GT}, the universe is a system like any other, governed by an equation of motion of the same form as the form of the equations of motion that govern its subsystems, the very subsystems whose dynamics are directly confirmed by experiments. The empirical success of \textbf{ST} thus itself lends as much \emph{prima facie} support to the view---the open systems view---according to which we should conceive of the universe's dynamically evolving state in terms of the dynamically evolving state of an open system, as it does to the closed systems view. The most general quantum-theoretical framework for describing the dynamically evolving state of an open system is \textbf{GT}.

The question of \emph{prima facie} empirical support is an epistemic question, and of course it is not the only such question that one can ask. In addition to epistemic questions, there is, further, the question of which of the two frameworks provides more fundamental explanations of phenomena. Motivated by this we now consider two alternative notions of fundamentality: epistemic fundamentality (Section \ref{s:epistemic-fund}) and explanatory fundamentality (Section \ref{s:explanatory-fund}).

\subsubsection{Epistemic Fundamentality}
\label{s:epistemic-fund}

In the previous subsection we argued that \textbf{GT}, formulated in accordance with the open systems view, is ontologically more fundamental than \textbf{ST}, formulated in accordance with the closed systems view. Moreover we saw that the reasons to embrace \textbf{GT} as an ontologically more fundamental theoretical framework stem from considering the ontology of none other than \textbf{ST}. We now turn to epistemic fundamentality and ask whether \textbf{GT} should also be thought of as epistemically more fundamental than \textbf{ST}. We will conclude that it should.

We begin, as before, by attempting to explicate what we mean when we say that one theoretical framework is epistemically more fundamental than another. Our first candidate explication draws on the notion of (epistemic) justification.\footnote{See, for instance, \citet[]{bennettbook}, who writes: ``The epistemically fundamental or foundational propositions $\dots$ are those that are not inferentially justified'' (p.\ 105). Stated this way, the condition actually expresses a sense of absolute epistemic fundamentality, but it is easy to extend it so as to capture the notion of relative epistemic fundamentality, as we do in \textbf{EpFund-1}.
} 
\bq
{\bf (EpFund-1)} Let ${\cal F}_F$ and ${\cal F}_P$ be two competing theoretical frameworks. Then ${\cal F}_F$ is \emph{epistemically more fundamental} than ${\cal F}_P$ iff ${\cal F}_F$ {\em justifies} ${\cal F}_P$ but not vice versa.
\eq
This explication presupposes that we understand what it means for ${\cal F}_F$ to justify ${\cal F}_P$. One way to flesh this out is to request that each theory $T^{(i)}_P$ in ${\cal F}_P$ reduce to some theory $T^{(j)}_F$ in ${\cal F}_F$, e.g.~in the sense of the Generalized Nagel-Schaffner (GNS) model of reduction \citep[see][]{GNS1,GNS2}. In this case one has to show (leaving some important details aside for the moment) that the laws of each $T^{(i)}_P$ can be derived from the laws of some $T^{(j)}_F$ and certain auxiliary assumptions.

Now we have already seen that \textbf{GT} is a more general framework than \textbf{ST} insofar as there are dynamical possibilities that one can make physical sense of in \textbf{GT} but not in \textbf{ST}. These relate to the dynamical evolution of the universe as a whole, which in \textbf{GT}, but not in \textbf{ST}, may be described by a not completely positive map. Since a completely positive map on the state space of a system is just a special case of a not completely positive map for which the compatibility domain of the map is defined on the system's entire state space, one can derive the dynamical laws of any theory formulated in \textbf{ST} from the dynamical laws of a corresponding theory formulated in \textbf{GT}, which seems to imply that \textbf{GT} is more fundamental than \textbf{ST} according to \textbf{EpFund-1}.

This conclusion would be premature, however. As we noted during our discussion of ontic fundamentality, appealing to the greater expressive power of \textbf{GT} in order to argue that \textbf{GT} is more fundamental than \textbf{ST} (in any sense) is circular unless one gives some independent motivation for taking the possibilities that one cannot make physical sense of in \textbf{ST} seriously. We provided this in the context of our argument for the ontological fundamentality of \textbf{GT}, but our intention in this section is to argue for its epistemic fundamentality independently of that argument. Thus let us restrict our discussion to only the possibilities that can be made physical sense of in both frameworks.

We have already seen that the equations associated with the framework of \textbf{GT} (in particular the Lindblad equation) can be derived from the equations of \textbf{ST} by embedding the open system under consideration into a larger closed system. At the same time we have also seen that one can derive the equations used in the framework of \textbf{ST} (e.g., the Schr\"odinger equation in the nonrelativistic case) from the equations of \textbf{GT} by simply disregarding the Lindblad equation's non-unitary terms (see Eq.\ \eqref{lindrhoc}). So, there seems to be a tie by the lights of \textbf{EpFund-1}: \textbf{GT} can be understood as justifying (the relevant equations of) \textbf{ST}, and the opposite is also true.

This is a little too quick, however. The GNS model of reduction, as stated so far, places no restrictions on the strength of one's auxiliary assumptions. In principle they could be used to do all of the actual reductive work, in which case the claim that ${\cal F}_F$ justifies ${\cal F}_P$ would seem to be implausible. To address this concern, \citet{GNS1} request two conditions (in the context of their discussion of intertheoretic reduction) which we adapt here: (i)~The {\em condition of non-redundancy} requests that ${\cal F}_P$ must not follow from the auxiliary assumptions alone. (ii)~The {\em condition of immanence} requires that the auxiliary assumptions must not be foreign to the conceptual apparatus of ${\cal F}_F$. They have to be a `natural part' of ${\cal F}_F$. 

Both conditions hold in our case study, both when it comes to justifying \textbf{ST} in terms of \textbf{GT} and vice versa. First, to derive the Schr\"odinger equation governing closed system evolution in the nonrelativistic version of \textbf{ST} (and likewise for the relativistic Dirac equation), one starts with the Lindblad equation and simply disregards the non-unitary terms (viz. the $L_i$ terms), which in any case make no contribution to the evolution of a closed system. Clearly, this violates neither condition~(i) nor~(ii). Second, we have already seen that the auxiliary assumptions that one has to make to derive the Lindblad equation (e.g. that the open system of interest is embedded in a larger closed system) are part of the modeling toolbox of \textbf{ST}. Therefore we conclude, once again, that neither \textbf{ST} nor \textbf{GT} is epistemically more fundamental than the other, at least by the lights of {\bf EpFund-1}.

We remark that we should not be surprised by this given that {\bf EpFund-1}, like \textbf{OntFund-1}, essentially concerns a set of logical relations obtaining between abstract concepts. Similarly to the way that we needed to consider the ontologies of the frameworks under consideration in order to settle the question of ontic fundamentality, giving a proper account of the relation of epistemic fundamentality between two given theoretical frameworks will require that we consider the way that their respective (`capital-O') Objects actually come to be known.

Accordingly, our second candidate explication of epistemic fundamentality draws on the idea that the objects of a given theoretical framework can be given an epistemic ordering \citep[cf.][sec.\ 64]{carnap1928}.\footnote{Although the notion of epistemic order we consider here is inspired by Carnap's discussion of the autopsychological basis in the \emph{Aufbau}, we depart from Carnap's exposition in many ways; for instance we do not require that a concept $A$ that occurs later than $B$ in the epistemic order be constructible from $A$ in terms of an explicit definition. One way that we do not depart from Carnap is that we also refer to abstract theoretical concepts as objects \citep[sec.\ 5]{carnap1928}.} An epistemic ordering of two given objects (of a given theoretical framework) relates them in terms of the actual order in which they come to be constructed in the context of an empirical investigation conducted using the modeling tools of the framework. Consider:
\bq
{\bf (EpistemicOrder)} Let ${\cal F}$ be a theoretical framework and $\mathcal{C}_1$ and $\mathcal{C}_2$ be two objects in ${\cal F}$. Then $\mathcal{C}_1$ comes before $\mathcal{C}_2$ in the epistemic order of objects of ${\cal F}$ iff any instance of $\mathcal{C}_2$ is always constructed on the basis of an instance of $\mathcal{C}_1$, but never vice versa, in the context of an empirical investigation being conducted in ${\cal F}$.
\eq

Let us apply this definition to the case of state vectors and density operators in \textbf{ST}.\footnote{For the time being we are restricting our attention to \textbf{ST} as the only way to model the state of a system in \textbf{GT} is in terms of a density operator.} 
We first consider the cases in which we assign a state vector to a system in the context of an empirical investigation of some phenomenon. There are essentially only two. The first corresponds to the situation in which we have designed some preparation device to prepare any number of systems of a particular type, e.g., silver atoms, in exactly the same way (i.e., so that it physically interacts identically with each system). In this case, assuming that the device is working perfectly, any system of the right type that interacts with it will be found to be in a particular state $| \psi \rangle_{\mathcal{S}}$ with certainty.

As we know, however, state preparation devices are never actually perfect. Thus strictly speaking we are never warranted in assigning the pure state $| \psi \rangle_{\mathcal{S}} \,_ {\mathcal{S}}\langle \psi |$ corresponding to the state vector $| \psi \rangle_{\mathcal{S}}$ to the ensemble generated by the device. Strictly speaking the ensemble will be in the mixed state:
\begin{align}
  \label{eqn:almost-pure}
\rho_{\mathcal{S}} = (1-\varepsilon) \,| \psi \rangle_{\mathcal{S}} \,_ {\mathcal{S}}\langle \psi | ~+~ \varepsilon \, \sigma_{\mathcal{S}},
\end{align}
where $\rho_{\mathcal{S}}$ and $\sigma_{\mathcal{S}}$ are density operators and $\varepsilon$ is some positive real number that (let us suppose) is very close to zero.

According to Gleason's theorem \citeyearpar{gleason1957}, quantum theory's assignment of probabilities is complete in the sense that every probability measure over yes-or-no questions concerning the observable quantities associated with a system is representable by means of a density operator acting on the system's state space,\footnote{The proof assumes that measurements are represented as projections and is valid for Hilbert spaces of dimension $\geq 3$.} but not, in general, by a state vector. If our level of confidence in the preparation device is high enough to render any small errors in its construction insignificant for the purposes of a given investigation, however, then we are permitted an idealization; i.e., we can disregard the information pertaining to the error term $\varepsilon$ in Eq.\ \eqref{eqn:almost-pure} and characterize the ensemble using the pure state $\rho_{\mathcal{S}} = | \psi \rangle_{\mathcal{S}} \,_ {\mathcal{S}}\langle \psi |$ corresponding to the state vector $| \psi \rangle_{\mathcal{S}}$ instead. The epistemic order in this case, when it comes to our characterization of an ensemble that has been prepared using our preparation device, is that we begin with the mixed state for $\mathcal{S}$ characterized by the density operator given in Eq.\ \eqref{eqn:almost-pure}, and then end (throwing some information away in the process) with an assignment of a pure state to $\mathcal{S}$ that we can describe using the state vector $| \psi \rangle_{\mathcal{S}}$.

The second case is where the goal is to describe the dynamics of an open system $\mathcal{S}$, whose state is described by a density operator $\rho_{\mathcal{S}}$, in interaction with its environment $\mathcal{E}$. As we saw in Section \ref{s:st}, to describe the dynamics of $\mathcal{S}$ we first assign the state vector, $| \Psi \rangle_{\mathcal{S+E}}$, to the combined system $\mathcal{S+E}$, unitarily evolve that state vector forward in time, and then trace over the degrees of freedom of $\mathcal{E}$. Note that no information concerning the system is thrown away in the process; thus assigning the state vector, $| \Psi \rangle_{\mathcal{S+E}}$, to the combined system $\mathcal{S+E}$ cannot be construed as an idealization (at least not in the same sense as our first case). All the same there can be no doubt in this case as well that, in terms of their epistemic order it is the density operator we assign to $\mathcal{S}$, rather than the state vector we assign to $\mathcal{S+E}$, that comes first.

Moreover, since (perfect) global measurements are not actually possible in \textbf{ST} \citep[see][]{vaidman2003},\footnote{The same goes for \textbf{GT}.} the only direct epistemic access that one can have to a system is through its subsystems. So if we want to learn something new about the combined system $\mathcal{S+E}$ we will have to do this by first interacting with one of $\mathcal{S}$ or $\mathcal{E}$, whose states are given by density operators.

Before moving on we should emphasize that (so far) we have \emph{not} argued that the concept of a density operator is \emph{explanatorily} more fundamental than the concept of a state vector (we will discuss explanatory fundamentality in Section \ref{s:explanatory-fund}). At this point we are only considering the evidentiary basis upon which a given state vector or density operator is assigned to a system. With regard to this we have just seen that in both cases where one assigns a state vector to a system, this is the endpoint of an empirical investigation into the system that begins with the assignment of a density operator.

Having discussed the notion of the epistemic order of objects in the context of a given framework, we can now present our second attempt to explicate what it means for one framework to be epistemically more fundamental than another.
\bq
{\bf (EpFund-2)} Let ${\cal F}_F$ and ${\cal F}_P$ be two competing theoretical frameworks. Then ${\cal F}_F$ is \emph{epistemically more fundamental} than ${\cal F}_P$ iff at least one object in ${\cal F}_F$ always comes before (and none after) a corresponding object in ${\cal F}_P$ in the epistemic order of objects.
\eq
In plain English, the upshot of \textbf{EpFund-2} is that ${\cal F}_F$ is epistemically more fundamental than ${\cal F}_P$ iff at least one object of ${\cal F}_P$ is constructed on the basis of a corresponding object (or a set of objects) of ${\cal F}_F$ and none vice versa. Note that the above explication of relative epistemic fundamentality assumes that ${\cal F}_F$ and ${\cal F}_P$ have some objects in common. This is true in our case study, i.e., they both include the density operator, to which we now return.

Density operators, as we have already argued, come before state vectors in the epistemic order of objects of \textbf{ST}. They also come first in \textbf{GT}, simply because state vectors do not appear at all in that framework.
Hence, there is at least one object of \textbf{ST} (viz. the state vector) that is constructed on the basis of a corresponding object (viz. the density operator) that is shared in common between \textbf{ST} and \textbf{GT}. Furthermore, there does not seem to be an object belonging to \textbf{GT} that is constructed on the basis of a corresponding object of \textbf{ST}. Hence, we conclude that \textbf{GT} is epistemically more fundamental, by the lights of \textbf{EpFund-2}, than \textbf{ST}. Since the alternative, \textbf{EpFund-1}, only captures the logical relations between the objects of two given frameworks, and does not consider how particular instances of those objects actually come to be known, \textbf{EpFund-2} is to be preferred as an explication of the relation of relative epistemic fundamentality over \textbf{EpFund-1}. We thus conclude that \textbf{GT} is epistemically more fundamental than \textbf{ST}.

Before we move on to the next section a disclaimer is in order. Although we have argued that some theoretical frameworks are epistemically more fundamental than others, we do not take ourselves to be committed to a foundationalist epistemology, for three reasons. First, foundationalist epistemologies generally appeal to relations of justification to argue that a given set of propositions is more fundamental than some other set. But we saw that, at least for our own case study, the more useful criterion for fleshing out the relation of epistemic fundamentality turned out to be that of epistemic order. Second, regardless of the criterion one uses to argue that some framework ${\cal F}_F$ is epistemically more fundamental than some other framework ${\cal F}_P$, it does not follow that all theoretical frameworks can be so ordered. Even in a coherentist epistemology it is quite possible for one set of statements to be more fundamental than another \citep{calosi2021}. Our own case study is somewhat special insofar as \textbf{GT} and \textbf{ST} are both theoretical frameworks for dealing with quantum phenomena and moreover share many core concepts in common. This is not always the case when one compares a given theoretical framework with another. The third reason has to do with our general argumentative strategy in this paper. None of our arguments for the ontic, epistemic, or explanatory fundamentality of \textbf{GT} appeal to the idea that quantum theory (as expressed by either \textbf{GT} or \textbf{ST}) should be thought of as a foundation for the rest of science. Instead, it is the success of the open systems view in the special sciences that motivates us in the first place to more closely investigate the supposed last refuge of the closed systems view: quantum theory.

\subsubsection{Explanatory Fundamentality}
\label{s:explanatory-fund}

We have argued that \textbf{GT} is epistemically and ontologically more fundamental than \textbf{ST}. There is, however, a further sense in which one of the two frameworks might be thought to be more fundamental than the other, viz. explanatory fundamentality. Unfortunately it is hard to get a grip on what, exactly, is meant by this, mainly because the notion of explanation is so elusive and has been understood so differently by different people. Thus in this subsection we will consider three possible ways to make sense of explanatory fundamentality in the light of our case study. We will argue that, {\em if} one wants to entertain the notion of `explanatory fundamentality' at all, then the right conclusion to draw, on the basis of our having determined that \textbf{GT} is ontologically and epistemically more fundamental than \textbf{ST}, is that it is explanatorily more fundamental than \textbf{ST} as well.

Some will find this conclusion implausible, or at least surprising, as \textbf{ST} often provides, or so one could argue, \emph{better} explanations than \textbf{GT}. That is to say, typically the explanations provided by \textbf{ST} are simpler and easier to grasp than explanations provided by \textbf{GT}; besides this, calculations involving closed systems are typically (though not always; see \citeauthor{hartmann2016}, \citeyear{hartmann2016}) more tractable than calculations that construe the systems under consideration as open; finally, the explanations that obtain from \textbf{ST} arguably provide us with a deeper {\em understanding} of a system, as they ignore noise which, at least on the closed systems view, one should regard as a disturbance that obscures what the actual difference-making factors that pertain to a system are.
These considerations would seem to suggest that the framework of \textbf{ST} is explanatorily more fundamental than \textbf{GT} rather than vice versa.

At this point it will be helpful to make the notion of explanatory fundamentality more precise. Accordingly we now present, and critically evaluate in the light of our case study, three candidate explications of relative explanatory fundamentality as it pertains to two competing theoretical frameworks. 

\bq 
{\bf (ExFund-1)} Let ${\cal F}_F$ and ${\cal F}_P$ be two competing theoretical frameworks. Then ${\cal F}_F$ is \emph{explanatorily more fundamental} than ${\cal F}_P$ iff ${\cal F}_F$ is \emph{more fundamental} than ${\cal F}_P$. 
\eq
The idea motivating {\bf ExFund-1} is simple: If we come to the conclusion that one theoretical framework is more fundamental than some other framework on some given criterion (which, presumably, has nothing to do with explanation \emph{per se}), then it follows that it is also explanatorily more fundamental than \textbf{ST} according to {\bf ExFund-1}. Of course, {\bf ExFund-1} presupposes that we have a criterion at hand which will allow us to decide whether one given framework is more fundamental than another given framework. This criterion might be the conjunction of ontic and epistemic fundamentality (as explicated in the previous subsections). But in that case explanatory fundamentality becomes fully dependent on ontic and epistemic fundamentality, and in particular it becomes impossible for ${\cal F}_F$ to be ontologically and epistemically more fundamental than ${\cal F}_P$ at the same time as ${\cal F}_P$ is explanatorily more fundamental than ${\cal F}_F$. Hence, if one concludes (as we have) that \textbf{GT} is ontologically and epistemically more fundamental than \textbf{ST}, then one also has to conclude that \textbf{ST} is explanatorily more fundamental than \textbf{ST}, at least according to \textbf{ExFund-1}. This may seem unsatisfying, as presumably we would like to get hold of a notion of explanatory fundamentality that is (at least to some extent) independent of ontic and epistemic fundamentality. Let us therefore move on to our next candidate explication.

\bq 
{\bf (ExFund-2)} Let ${\cal F}_F$ and ${\cal F}_P$ be two competing theoretical frameworks. Then ${\cal F}_F$ is \emph{explanatorily more fundamental} than ${\cal F}_P$ iff ${\cal F}_F$ \emph{explains} ${\cal F}_P$ but not vice versa.
\eq 

This explication presupposes that we know what it means for one framework to explain another framework. For instance, we might flesh out this notion analogously to the way we fleshed out (in Section \ref{s:epistemic-fund}) the notion of justification that appears in the explication of {\bf EpFund-1}, arguing that ${\cal F}_F$ explains ${\cal F}_P$ iff each theory in ${\cal F}_P$ reduces to a theory in ${\cal F}_F$ according to the GNS model of reduction. But while this could perhaps be done, it is doubtful in any case whether {\bf ExFund-2} can really capture what is meant by explanatory fundamentality as it does not consider the actual explanations that ${\cal F}_F$ and ${\cal F}_P$ provide of phenomena. It only considers an abstract formal relation that holds between the theories of two given frameworks.

Let us elaborate on this idea. Assume, first of all, that each theoretical framework under consideration sets out to account for a certain set of phenomena, noting that phenomena can be (and in general are) mathematized and mediated by theory.\footnote{See \cite{bogenwoodward1988, harper2011, woodward2011}.} For instance John Bell's analyses \citeyearpar[]{bell1964, bell1966} of the differences between quantum and local hidden-variable theories are carried out in relation to probabilistic phenomena that are analyzable using the tools of elementary probability theory. Although there are differences in the ways that probability statements can be assigned in local hidden-variable theories and in quantum theory, respectively \citep[]{dickson2011,pitowsky1989}, we can nevertheless talk about probabilities in both cases. Probabilistic phenomena thus constitute a shared domain.

Let us also assume that there is a considerable amount of agreement with respect to the set of (possible or actual) phenomena in the domain in question that can be accounted for in both frameworks. Of course, this does not necessarily mean that the sets of phenomena that can be accounted for in both frameworks are identical. It is perfectly conceivable that the set of phenomena that can be given an account of in one framework is a proper superset of the set of phenomena that can be given an account of in another framework.\footnote{In our own case study of \textbf{GT} and \textbf{ST}, two theoretical frameworks for describing probabilistic quantum phenomena, we have seen that \textbf{GT} describes possible dynamical evolutions that cannot be made physical sense of in \textbf{ST}. No phenomena are known that would absolutely force us to consider one of these dynamical possibilities, but it is well known that certain phenomena suggest that we should take them seriously (we will come back to this in Section \ref{s:phil-impl}). 
} But we can still ask, regarding the phenomena that they both are able to give us an account of, whether one or the other theoretical framework gives us a better account, where by `better' we mean (as we mentioned above) an account that is simpler, more tractable, and easier to understand. These considerations suggest the following explication.
\bq 
{\bf (ExFund-3)} Let ${\cal F}_F$ and ${\cal F}_P$ be two competing theoretical frameworks such that ${\cal F}_F$ accounts for a set ${\cal P}_F$ of phenomena and ${\cal F}_P$ accounts for a set ${\cal P}_P$ of phenomena with ${\cal P}_P \subseteq {\cal P}_F$. Then ${\cal F}_F$ is \emph{explanatorily more fundamental} than ${\cal F}_P$ iff ${\cal F}_F$ explains the common phenomena ${\cal P} = {\cal P}_P$ \emph{better than} ${\cal F}_P$.
\eq 

Note that on this explication, ${\cal F}_F$ is explanatorily more fundamental than ${\cal F}_P$ relative to a set of phenomena ${\cal P}$ that both frameworks are able to successfully explain. 
To decide which of the corresponding frameworks in question is explanatorily more fundamental according to {\bf ExFund-3}, we therefore have to investigate which framework provides the better explanations (in the sense described above) of that common set of phenomena.

Recall the remarks that we made at the beginning of this subsection. There we noted that \textbf{ST}'s explanations are typically simpler, more tractable, and easier to understand---i.e., better---than the explanations provided by \textbf{GT}. Hence, \textbf{ST} seems to be explanatorily more fundamental than \textbf{GT} according to {\bf ExFund-3}. Surely this cannot be right, however. Consider the following parallel argument: The framework of Newtonian mechanics provides a perfectly acceptable explanation of the orbit of Halley's comet, and why it is visible from Earth every 75--76 years, that is reasonably simple, tractable, and easy to understand. And yet, it is uncontroversial that general relativity is clearly and on all counts the more fundamental theoretical framework, and it seems reasonable, on this basis, to also insist that general relativity provides more fundamental explanations of orbital phenomena than Newtonian mechanics. Since {\bf ExFund-3} will not allow us to conclude that general relativity is explanatorily more fundamental than Newtonian mechanics, and since we take that to be the wrong answer, we therefore conclude (by \emph{reductio ad absurdum}) that explanatory goodness does not entail explanatory fundamentality, at least not in science. In science, the more fundamental explanation is not necessarily the better explanation (as perhaps \citep{weinbergdreams} thought in {\em Dreams of a Final Theory}).\footnote{It is interesting to note that explanatory considerations play an important role in discussions of fundamentality in metaphysics. Our discussion sheds a critical light on this: Explanatory goodness is not always a reliable guide to fundamentality.} 

It is time to take stock again and see where we stand. In this subsection, we have (i)~criticized {\bf ExFund-1} as deflationary, (ii)~arrived at the conclusion that there is a tie according to {\bf ExFund-2}, which we dismissed as not really tracking what we mean by explanatory fundamentality in any case, and (iii)~argued---against {\bf ExFund-3}---that explanatory betterness is not a reliable guide to relative explanatory fundamentality. This suggests that {\em if} one wants to entertain the notion of `explanatory fundamentality' at all, then it is best to use the deflationary explication (that is, {\bf ExFund-1}) and argue that the explanatorily more fundamental framework is the framework that is fundamental as determined by some independent criterion. For the present case study, since we have already concluded that \textbf{GT} is both ontologically and epistemically more fundamental than \textbf{ST}, we should conclude that \textbf{GT} is explanatorily more fundamental than \textbf{ST} as well.

However, if one sees no value in accepting the deflationary explication {\bf ExFund-1}, then one can discard the notion of `explanatory fundamentality' altogether. For while explanatory goodness may be an important scientific virtue, it is unlikely to be a reliable guide to fundamentality as well. In any case this is not so in science. Therefore, talk of explanatory fundamentality should be replaced, we think, by talk of explanatory goodness, and the notion of `explanatory fundamentality' is probably best discarded altogether.

\subsection{Fundamentality \emph{tout court}}
\label{s:tout-court}

We have now argued that \textbf{GT}, a theoretical framework for describing quantum phenomena formulated in accordance with the open systems view, is more fundamental than \textbf{ST}, a theoretical framework for describing quantum phenomena formulated in accordance with the closed systems view, regardless of whether one cashes out fundamentality in ontic, epistemic, or explanatory terms. We take it that this is enough to conclude that the open systems view is more fundamental than the closed systems view in the context of quantum theory. Quantum theory was, as we mentioned in Section \ref{s:intro}, the last best hope of the closed systems view in the following sense. Although, as we saw in Section \ref{s:intro}, the question of whether to admit the possibility of fundamental non-unitary evolution in gravitational physics is controversial, one of the most important reasons to resist making such a move arises from the desire to remain consistent with quantum theory. Once it is concluded that the open systems view is fundamental in quantum theory, however, these considerations are moot. We take there to be no reason in principle, therefore, to bar the researcher from considering the fundamental dynamics of any system in terms of the dynamics of an open system. Note further that the open systems view is compatible with the conclusion that the universe is a closed unitarily evolving system if that turns out to be the best way to describe it. As we have mentioned already, the open systems view does not deny that closed systems exist. It merely denies that \emph{only} closed systems exist, and also that a closed system \emph{must} exist, and represents \emph{all} systems as, in general, open. We take it that this is enough to conclude that the open systems view is more fundamental than the closed systems view throughout science. And since there are no other options, we conclude that the open systems view is fundamental \emph{tout court}.


\section{Wider Implications}
\label{s:implications}

Having argued, in the previous section, that the open systems view is fundamental; in this section we discuss a number of further implications of the open systems view for the foundations and philosophy of physics (Section \ref{s:phil-phys-impl}), the philosophy of science (Section \ref{s:phil-sci-impl}), and for metaphysics (Section \ref{s:phil-impl}).

\subsection{Foundations and Philosophy of Physics}
\label{s:phil-phys-impl}

Recall from Section \ref{s:views-frameworks-qt} that the closed systems view, in the context of \textbf{ST}, is cashed out in terms of the assumption that the dynamical evolution of the state of a closed system is unitary. As we pointed out in Section \ref{s:intro}, arguably the only closed system that truly exists, and thus the only system that can truly be described in these terms, is the universe as a whole. To begin with we note that it is not even clear what such a picture really amounts to. One of the more promising approaches to quantizing gravity, for instance, is the so-called quantum reference frames approach \citep[see, e.g.][]{castroEtAl2020, giacominiEtAl2019}, where, rather than modeling time as an external parameter, one treats a reference clock as a quantum system like any other. Although relative to a given clock one can describe the evolution of a given system in unitary terms, no single such clock is preferred, so that the unitary evolution one describes will be different depending on the reference clock chosen. This is clearly \emph{not} the simple picture of an isolated quantum system to which one objectively assigns a particular state vector unitarily evolving through time \citep[see also][sec.\ 3]{oritiTimeEmergence}.

Further conceptual problems arise for the closed systems view when we consider the physics of black holes and the so-called problem of the arrow of time. As we alluded to in Section \ref{s:intro}, in a series of papers in the 1970s, Stephen Hawking \citeyearpar[]{hawking1975, hawking1976a, hawking1976b} showed that a black hole will radiate particles over time until it shrinks to a mass comparable to the Planck mass, at which point current theories of physics provide no sensible answer to the question of what happens next. Hawking's original idea \citeyearpar[]{hawking1976b} was that after reaching the Planck mass a black hole then disappears in a final burst. The conceptual problem with this proposal, however, is that it entails that, post-evaporation, the state of the radiated particles will be mixed, despite the fact that they are no longer entangled with any corresponding particles in the black hole (which will have since disappeared). This constitutes a \emph{prima facie} violation of unitarity, and thus a \emph{prima facie} conflict with \textbf{ST}. Alternative accounts of the final stages of black hole evaporation have been proposed, but these arguably involve even greater conceptual problems.\footnote{For reviews and for further discussion, see \citet[]{giddings2013}, \citet[]{huggettMatsubara2021}, \citet{maudlinInfoLost}, and \citet[]{wallace2020a}.} The problem of the arrow of time \citep[]{callendarTimeAsymmetrySEP}, relatedly, is essentially the problem of how to reconcile the fact that, at the phenomenological level, physical processes generally display temporally asymmetric behavior, with the fact that physics (both in classical mechanics and in its successor, \textbf{ST}) describes the dynamics of systems as reversible, which in the case of \textbf{ST} follows from the fact that the evolution of a state vector is unitary.

On the open systems view, however, none of these basic conceptual problems arise. It is not a problem that the complementary pictures of the unitary evolution of the universe one obtains from the quantum reference frames approach are perspective-dependent, since that is only a problem if one insists that closed systems are fundamental. As for the information loss paradox: Although we do not want to pronounce on the correct answer to the question of what happens to a black hole as its mass approaches the Planck scale size, we think it is important to point out that Hawking's original proposal to describe the dynamics of radiated particles in terms of linear non-unitary maps on density matrices \citep[cf.][sec.\ V]{hawking1976b}, despite its apparent conflict with \textbf{ST}, is conceptually unproblematic since this is, as we saw, consistent with the allowed dynamical evolutions of a quantum system given by \textbf{GT}. The case of the arrow of time is similar: Although we leave it to others to work out the details,\footnote{See, for instance, \citet[][]{chenTimeAsymmetricBJPS, chenNatureOfPhysicalLaws, chen2020}.
Chen also adds (but does not require) a special initial state of the universe---a generalization of David Albert's \citeyearpar[]{albert2000} `past hypothesis'---which he calls the ``initial projection hypothesis.'' Owen Maroney (who does not posit a special initial state) has expressed similar ideas both in informal correspondence and in presentations on the topic over the years. See also \citet{maroneyGibbsPhysicalBasis} and \citet{robertsonHolyGrail} for related discussions of how taking the density matrix seriously as a real representation of the state of a system helps to clarify the physical basis of the Gibbs entropy and its role in providing a statistical-mechanical underpinning for the second law of thermodynamics.} the basic conceptual problem disappears in a framework in which the allowed dynamical evolutions of a quantum system may be non-unitary.

\subsection{Philosophy of Science}
\label{s:phil-sci-impl}

In Section \ref{s:views-frameworks-qt} we explicated the concept of a view as representing an approach to science that is associated with (i)~a set of methodological presuppositions in accordance with which we characterize the objects of a given domain, that is (ii)~motivated by a particular metaphysical position concerning the fundamental nature of the objects that one aims to characterize. It is interesting to note that a view, construed in this way, has elements in common with the concept of a {\em stance} as developed by Bas \citet{vanFraassen2002}. As van Fraassen explains, a stance is ``not identifiable with a theory about what there is but only with an attitude or cluster of attitudes'' (ibid., p. 59). Thus a stance, like a view, is associated with a particular methodology (i.e., the `attitudes' just mentioned). But unlike a view, a stance does not (or at least it is not intended to) refer to a motivating metaphysical position. This is of course very natural given van Fraassen's anti-metaphysical tendencies (ibid., Lecture 1).

Recall from Section \ref{s:ontic-fund} that the fact that a view is motivated by a particular metaphysical position concerning the fundamental nature of the objects of a given domain is what allowed us to argue for the relative fundamentality of \textbf{GT} to \textbf{ST} in the context of quantum theory; for we saw that despite being formulated in accordance with the closed systems view, which is motivated by the metaphysical position that takes closed systems to be fundamental, carefully considering the way that \textbf{ST}'s (capital-O) Objects represent the (little-o) objects of its domain shows us that those objects are actually open systems either exclusively (on an orthodox interpretation) or in addition to closed systems (on an Everettian interpretation). Such an argument would clearly not be available in the case of a stance. While we take this to be an advantage of the concept of a view over that of a stance (i.e., while we take the concept of a view to be more useful for science), in fairness to van Fraassen we note that it was never his intention, in introducing the concept of a stance, that we should actually be able to rationally argue for one. Developing the concept of a stance as it applies to his specific take on empiricism, for instance, van Fraassen writes:
\begin{quote}
Being or becoming an empiricist will then be similar or analogous to conversion to a cause, a religion, an ideology, to capitalism or to socialism, to a worldview such as Dawkins's selfish gene or the view Russell expressed in ``Why I Am Not a Christian''. (p.~61)
\end{quote}
Thus, a stance contains (and is intended to contain) an essential voluntaristic element. As presented by van Fraassen, it is also vague (or anyway has been criticized as such). Other authors \citep[see, e.g.,][]{boucher2014, chakravartty2011, lipton2004, rowbottom2011, teller2004} have attempted to make the concept more precise, in particular by appealing to pragmatic and value-based considerations to flesh out how one can choose between competing stances, but the essentially voluntaristic nature of a stance is nevertheless maintained by these authors.

As for a view, we have seen that it is not only possible to argue for a view in the context of a choice between two competing theoretical frameworks (e.g., \textbf{ST} and \textbf{GT}) characterizing a particular domain of inquiry. It is also possible to argue globally for a view, as indeed we did in this paper; i.e., we argued that since the open systems view is more fundamental than the closed systems view throughout existing science, that it is more fundamental than the closed systems view in general, and thus it is the view from which new theoretical frameworks should be formulated going forward. In other words we have rationally argued that it is time to abandon, as a criterion for the validity of a given theory or theoretical framework, that it be consistent with either the methodological or metaphysical presuppositions of the closed systems view.

\subsection{Metaphysics}
\label{s:phil-impl}

Recall, from Section \ref{s:views-frameworks-qt}, that associated with a given view is (i)~a set of methodological presuppositions in accordance with which we characterize the objects of a given domain, that is (ii)~motivated by a particular metaphysical position concerning the nature of those objects (and objects in general). Frameworks are formulated in accordance with a given view insofar as they formalize that view's methodological presuppositions and apply them universally to all the phenomena describable in the framework. As we discussed, \textbf{GT} is formulated in accordance with the open systems view, and this view is motivated by the metaphysical position that takes open systems---i.e., systems interacting with their environment---to be fundamental. \textbf{GT} represents open systems using density operators, and the fundamental equations of motion of \textbf{GT} describe the ways in which density operators may evolve over time into other density operators.

We have seen above that the subject matter of a theoretical framework---what a philosophical analysis of its objects leads us to conclude about the ontology they represent---does not necessarily conform to the metaphysical position that motivates us to formalize the framework in terms of those objects in the first place (see Section \ref{s:ontic-fund}). In the case of \textbf{ST} we saw that the metaphysical position that motivates the use of the state vector to represent quantum systems does not align with the way that it actually represents those systems: Although \textbf{ST} is formulated in accordance with the closed systems view, the systems it represents are in general open.

\textbf{GT} is formulated in accordance with the open systems view, which is less restrictive than the closed systems view insofar as it is consistent with the metaphysical position that motivates the former view that closed systems can exist. The obvious candidate for a closed system in \textbf{GT}, given that (let us presume) everything is in interaction with everything else, is the cosmos as a whole. One of the possible ways in which the cosmos can evolve in \textbf{GT} is unitarily, in which case interpreting it to be a closed system would seem to be conceptually unproblematic. However, \textbf{GT} does not require us to model the dynamical evolution of the cosmos as unitary---unlike \textbf{ST}, it allows for the dynamical evolution of the cosmos to be non-unitary. But what can this really mean? Should we interpret the non-unitary evolution of the cosmos as the dynamical evolution of an open or of a closed system? Neither concept would seem to be adequate. On the one hand, it would seem to make little sense to describe a non-unitarily evolving system as described by \textbf{GT} as closed. The generator of the quantum dynamical semigroup for such a system, as we saw in Section \ref{s:views-frameworks-qt}, includes terms besides those associated with the system's Hamiltonian, and these extra terms are most naturally interpreted as due to the system's interaction with an external system. It is even less clear how to interpret the evolution of the cosmos in the case where this is described in terms of a not completely positive map on its state space. Essentially such a map is a partial function on that state space, and it would be hard to defend the position that such a restriction---on the otherwise valid possible states of a system---should be included in our metaphysical conception of what that system is. On the other hand, it also seems to make little sense to say that the universe as a whole is an open system, at least not in the sense that we usually mean by this term. After all, the universe just means everything that exists, does it not?

It would seem, then, that the ontological distinction between open and closed systems (at least when it pertains to the cosmos) breaks down in \textbf{GT}. This is a lesson which metaphysicians should take to heart. Ultimately, density operators, state vectors, and other formal concepts in physics are bits of mathematics that we use to describe the world. It must not be presumed that every such concept that arises in a scientific theory will conform, when interpreted ontologically, to our pre-theoretic concepts or even to the concepts of our predecessor theories. Ultimately metaphysics and science must work together: Science provides the opportunity for metaphysics to re-conceptualize basic ontological distinctions such as that between an open and a closed system, while metaphysics, on the basis of this renewed understanding of the world, motivates new science.

What, then, can it mean to represent the cosmos as a non-unitarily evolving density operator? We will leave it for another occasion to fully explore this question, but we will gesture at some possible answers here: Given that (a) the dynamics of a non-unitarily evolving universe as described by \textbf{GT} would appear to us \emph{as if} it were the dynamics of a system interacting with an external system (i.e., it can be described as though it were the dynamics of a subsystem of an entangled system), and given that (b) the most general dynamical evolution describable in \textbf{GT} requires that we conceive of that larger system as including an unphysical component (see Section \ref{s:wider-framework}),\footnote{Recall that a not completely positive map on a system's state space can be understood as a contraction from the dynamics of a larger closed system only if we generalize our representation of the state space of the environment so that the combined state of $\mathcal{S+E}$ is no longer physically interpretable.} one might be tempted to conceive of the cosmos in the way that Newton did, i.e., as occasionally subject to (and indeed requiring) interventions from a domain of existence that is not described by physics. While this option is perhaps always open, it is not entirely satisfactory as it seems to us that the main motivation for understanding the universe in this way (coming from physics, at any rate) stems from the closed systems view.

An alternative, which is arguably more fully in the spirit of the open systems view, is to think of the form taken by the dynamical evolution of the cosmos as just a brute fact, which we can describe in \textbf{GT}, but which, like the principle of inertia in Newtonian mechanics, is not subject to further explanation. But just like the brute fact of the principle of inertia in Newtonian mechanics, the brute fact of the form of the dynamics of the universe is one that may potentially lead the way to new physics. Could this new physics lead us back to a kind of closed systems view in the end? We think this is unlikely, and anyway it would require a re-conceptualization of what a `closed system' is. In any case, while it would be a mistake to rule this out as a logical impossibility, we can certainly say that from our current standpoint, the open systems view, as we have described it in this paper, is fundamental in science. Wherever science leads us in the future, we will have arrived there by following the path laid out by it, not the path laid out by the view of science it has superseded. Metaphysicians are best served, therefore, in their discussions of the nature of the cosmos, in focusing on the description of it provided by the open systems view.


\section{Summary and Conclusion}
\label{s:conclusions}

In Section \ref{s:views-frameworks-qt}, we clarified (among other things) the important concepts of a theoretical framework and a view, as well as the two theoretical frameworks for characterizing quantum phenomena comprising our case study: \textbf{ST} and \textbf{GT}, after which we considered their respective interpretations in Section \ref{s:interpreting-st-and-gt}. In Section \ref{s:fundamentality} we then argued that \textbf{GT} is ontologically, epistemically, and explanatorily more fundamental than \textbf{ST}, and since the open systems view from which \textbf{GT} is formulated is more fundamental than the closed systems view from which \textbf{ST} is formulated in every other area of science,\footnote{We do not deny that in some cases it might be useful to construe the systems one is considering in a given domain as being closed, but as we mentioned earlier we take this to be the exception rather than the rule.} it follows that the open systems view is fundamental \emph{tout court}. Finally, in the last section we saw that taking the open systems view to be fundamental has important implications for the foundations and philosophy of physics, the philosophy of science, and for metaphysics. We close our discussion by speculating on some further foundational and philosophical issues that we think may be usefully informed by taking the open systems view to be fundamental.

\begin{enumerate}
\item \emph{Laws of nature:} We think it is likely that accepting that the open systems view is fundamental will have repercussions for the philosophical debate regarding fundamental laws of nature. In particular taking the open systems view to be fundamental entails that laws, such as the Dirac equation, associated with theories formulated in 
  \textbf{ST} can no longer be construed as fundamental and that we should formulate alternatives in the framework of \textbf{GT}. A natural starting point in the context of quantum field theory, perhaps, is the Schwinger-Keldysh formalism as it covers interactions that occur in open systems \citep{Haehl2017SchwingerKeldyshFP,jana2020}.
\item \emph{Theoretical equivalence:} we briefly discussed theoretical equivalence at the level of frameworks above, where we noted that \textbf{GT} has more expressive power than \textbf{ST}. But it is also worth investigating the relation of theoretical equivalence in the context of a comparison between different theories formulated in these frameworks (e.g., between relativistic theories formulated in \textbf{ST} vs. relativistic theories formulated in \textbf{GT}).
\item \emph{Levels of descriptions:} It would be interesting to give a detailed comparison between closed and open systems frameworks in a way that relates to the philosophical debate about levels of descriptions, levels of explanations, and ontological levels \citep{list2019}.
\item \emph{Extrinsic vs. intrinsic properties:} The philosophical position called \emph{ontic structural realism} (OSR) seems to rely on a distinction between extrinsic and intrinsic properties \citep[p. 151]{ladymanRoss2007}, but on the open systems view this distinction would appear to break down (recall that on the open systems view, the dynamical evolution of a system is not construed as arising from the system of interest's relation to other systems but is represented directly in the dynamical laws that govern the system). This raises the question of whether advocates of OSR really need to rely on the distinction between extrinsic and intrinsic properties in the first place, and whether OSR is consistent with the open systems view in light of this (we speculate that it is).
\item \emph{Causality:} Classical accounts of interventionist causality \citep[]{pearl2009, woodward2003}, as well as quantum generalizations \citep[]{allen2017, costaShrapnel2016, evans2018, evans2021, shrapnel2019}, rely essentially on the concept of an open system yet assume the closed systems view of physics (in quantum generalizations, they assume \textbf{ST}). As a consequence, some proponents (see \citealt{woodward2007} and for discussion and criticism see \citealt[]{reutlinger2013}) argue that interventionist accounts are not applicable to physics, at least not at the global scale \citep[sec.\ 6]{woodward2007}. Given that the universe is in principle describable as an open system in \textbf{GT}, it would thus be interesting to reconsider interventionist causality in that context.
\item \emph{Agency and free will:} \citet[]{briegel2015} have argued that the concept of an agent, from the point of view of quantum physics, strictly speaking corresponds to an open system that is in continuous interaction with its environment (p.\ 277), and have drawn on this fact to argue for a libertarian conception of agency and free will in the context of an indeterministic world. It would be interesting in this context to consider the more general framework of \textbf{GT}.
\item \emph{Persistence:} It seems relevant to the question of the persistence of an object through time whether that object is fundamentally construed as a closed system or as an in general open system interacting with its environment. Discussions of persistence that relate the issue to quantum mechanics, however \citep[see, e.g.,][]{pashby2016}, are generally conducted in the context of \textbf{ST}, which models the dynamics of open systems as only effective.
\end{enumerate}

These are only a few examples of questions in physics, the philosophy of science, and metaphysics that can be illuminated by taking the open systems view. The philosophy of open systems is only beginning!

\bigskip\bigskip

\noindent {\bf Acknowledgments:} We want to thank the members of our research group on the philosophy of open systems: James Ladyman, S\'ebastien Rivat, David Sloan, and Karim Th\'ebault; as well as the following people for discussion and in some cases for their critical comments on previous drafts of this paper: Jeffrey Bub, Jeremy Butterfield, Eddy Keming Chen, Erik Curiel, Richard Dawid, Gemma De les Coves, Neil Dewar, John Dougherty, Armond Duwell, Peter Evans, Sam Fletcher, G\'abor Hofer-Szab\'o, Johannes Kleiner, Kerry McKenzie, Markus M\"uller, Wayne Myrvold, Daniele Oriti, Alex Reutlinger, Katie Robertson, Simon Saunders, Jonathan Simon, Paul Teller, Frida Trotter, Lev Vaidman, David Wallace, and Naftali Weinberger. We also want to thank, for their helpful feedback, those in attendance at our presentations of preliminary versions of this paper at the Workshop on Symmetries and Asymmetries in Physics at Leibniz University Hannover, the British Society for the Philosophy of Science meeting in Oxford, the Philosophy of Science Association meeting in Seattle, the Work in Progress Seminar of the Munich Center for Mathematical Philosophy, the Canadian Society for History and Philosophy of Science online meeting hosted by the University of Alberta, the online Workshop on Experiment and Theory hosted by the University of Montreal, the Bristol Centre for Science and Philosophy Colloquium, the New Foundations Colloquium of LMU's Center for Advanced Studies, the conference: The Quantum, the Thermal and the Gravitational Reconciled at the Munich Center for Mathematical Philosophy, the Workshop on the Many Worlds Interpretation of Quantum Mechanics at Tel Aviv University, and the Workshop on Principles in Physics at the University of Wuppertal. Finally we thank the German Research Council (DFG), through grant number 468374455, and the Alexander von Humboldt Foundation, through an Experienced Researcher Grant to MEC, for their generous support.

\bibliographystyle{../apa-good}
\bibliography{../open_systems}{}

\end{document}